\documentclass[universe,article,accept,pdftex,moreauthors]{Definitions/mdpi}
\usepackage{graphicx}
\usepackage{amsmath,amssymb,amsfonts}
\usepackage{multirow}
\usepackage[utf8]{inputenc}
\usepackage{calc}
\usepackage{placeins}
\newcommand{\apj}{ Astrophys. J.}
\newcommand{\apjl}{ Astrophys. J. Lett.}

\newcommand{\aj}{ Astron. J.}
\newcommand{\aap}{Astron. Astrophys.}
\newcommand{\araa}{Annu. Rev. Astron. Astrophys.}
\newcommand{\mnras}{Mon. Not. R. Astron. Soc.}
\newcommand{\nat}{Nature}

\providecommand{\prl}{Phys.~Rev.~Lett.} 
\providecommand{\aapr}{ Astron. Astrophys. Rev.}
\DeclareUnicodeCharacter{2299}{\ensuremath{\odot}}
\providecommand{\pasa}{Publ. Astron. Soc. Aust.}

\newcommand{\kms}{\mathrm{km\,s^{-1}}}
\newcommand{\micron}{\ensuremath{\upmu\mathrm{m}}}

\newcommand{\Msun}{\ensuremath{\mathrm{\rm M}_\odot}}
\newcommand{\Mbh}{\ensuremath{M_\mathrm{\rm BH}}}
\newcommand{\Mnsc}{$M_\mathrm{\rm NSC}$}

\newcommand{\Lsun}{\ensuremath{\mathrm{\rm L}_\odot}}

\newcommand{\ml}{\ensuremath{M/L}}

\newcommand{\farcs}{^{\prime\prime}}

\newcommand{\arcsec}{^{\prime\prime}}

\mathchardef\mhyphen="2D

\renewcommand{\farcs}{\mbox{$.\!\!^{\prime\prime}$}}
\graphicspath{{./}{figures/}}

\firstpage{1} 
\pubvolume{12}
\issuenum{6}
\articlenumber{160}
\pubyear{2026}
\copyrightyear{2026}
\externaleditor{Yongquan Xue} 
\datereceived{15 April 2026} 
\daterevised{18 May 2026} 
\dateaccepted{22 May 2026} 
\datepublished{29 May 2026} 
\hreflink{https://doi.org/10.3390/universe12060160}

\Title{Probing the Variation of the Inner Surface-Brightness 
	Profile of~Nuclear Star Clusters on the Intermediate-Mass Black Hole Mass Measurements Using Mock Observations of ELT/MICADO and HARMONI}

\Author{
{Tinh Q. T. Le} 
 $^{1,}$*\orcidA{},
Dieu D. Nguyen $^{2}$\orcidB{},
Hai N. Ngo $^{2}$\orcidC{},
Tien H. T. Ho $^{3}$\orcidD{},
Tuan N. Le $^{3}$\orcidE{} and
Long Q. T. Nguyen $^{3}$\orcidF{}
}

\AuthorNames{Tinh Q. T. Le, Dieu D. Nguyen, Hai N. Ngo, Tien H. T. Ho, Tuan N. Le and Long Q. T. Nguyen}

\address{%
$^{1}$\quad Department of Physics, International {University,} 
 Vietnam National {University} in Ho Chi Minh City, \mbox{{Ho Chi Minh City 700000,} 
 Vietnam}\\
$^{2}$\quad Department of Astronomy, University of Michigan, 1085 South University Avenue, Ann Arbor, MI 48109, USA; {dieun@umich.edu (D.D.N.); hainn@umich.edu (H.N.N.)} 
\\
$^{3}$\quad Faculty of Physics--Engineering Physics, {University} 
 of Science, Vietnam National {University} in Ho Chi Minh City, {Ho Chi Minh City 700000,} Vietnam; {22130179@student.hcmus.edu.vn (T.H.T.H.); 25c3101344@student.hcmus.edu.vn (T.N.L.); 23130028@student.hcmus.edu.vn (L.Q.T.N.)}\\
}

\corres{Correspondence: {lethongquoctinh01@gmail.com} 
}

\abstract{
Simulations of intermediate-mass black holes (IMBHs) in dwarf galaxies within 10 Mpc that host bright nuclear star clusters (NSCs), prime candidates for IMBH formation, using the High Angular Resolution Monolithic Optical and Near-infrared Integral (HARMONI) field spectrograph on the Extremely Large Telescope, probe black hole formation in the early universe. Our approach combines observed surface-brightness profiles from the Hubble 
 Space Telescope (HST), synthetic stellar population spectra, and~Jeans Anisotropic Modeling (JAM) for stellar dynamics. Mock HARMONI observations were generated with the HSIM simulator and analyzed in a Bayesian framework to infer IMBH masses down to 0.5\% of the NSC mass. In~this work, we extend these simulations by constructing improved stellar mass models using SimCADO to simulate imaging with the Multi-AO Imaging Camera for Deep Observations (MICADO). The~MICADO data are jointly analyzed with HARMONI kinematics via JAM to reassess IMBH masses and uncertainties. This combined framework enables us to examine how variations in the NSC inner surface-brightness slope influence IMBH mass estimates, providing tighter constraints on low-mass black holes and advancing models for IMBH detection in~NSCs.    
}


\begin{document}
\section{Introduction}\label{intro} 

There is a fascinating, but~elusive, population of intermediate-mass black holes (IMBHs, {$M_{\rm BH} \approx 10^{2} - 10^{5}$} 
 \Msun) in low-mass galaxies {[$M_{\star}<10^{10}$ \Msun];~\citep{Ahn18, Voggel18, Nguyen19conf, Thater2022, Thater23, Zou2025},} 
 falling between the masses formed by single-massive stars today ($M_{\rm star} > 100$ \Msun) and supermassive black holes [SMBHs, $M_{\rm BH} \approx 10^{6-10}$ \Msun];~\citep{vandenBosch2016, Nguyen23, Nguyen2025a, Nguyen2025c, Nguyen2025d}, which are often found at the centers of massive galaxies [$M_{\star}\gtrsim10^{10}-10^{12}$ \Msun];~\citep{McConnell13, Saglia16}. The~compelling evidence of IMBHs was the LIGO/Virgo discovery of an IMBH formed from a collision of two smaller black holes [BHs, $M_{\rm BH}=142^{+28}_{-16}$ \Msun];~\citep{Abbott20}. However, current ground-based gravitational wave interferometers are limited to detecting black holes (BHs) with $M_{\rm BH} < 200$ \Msun. For~ $M_{\rm BH}\approx10^{3-5}$ \Msun, the~search must be carried out with integral field spectroscopic (IFS) observations in nearby stellar systems [{hereafter} 
 {N25}]~\citep{Nguyen2025b} for stellar dynamics with the 39-meter European Extremely Large Telescope (ELT) and its High Angular Resolution Monolithic Optical and Near-infrared Integral (HARMONI) field spectrograph instrument \citep{Thatte24}.   

It is also not yet clear if there is a full spectrum of IMBH masses \citep{Mezcua17}. Nonetheless, understanding the abundance, frequency, and~mass distribution of IMBHs contributes significantly to our cosmic inventory \citep{Fukugita04}, providing fundamental insights into the structure of our universe. Previous studies have profound implications for understanding how SMBHs formed in terms of IMBH seeds in the early universe through several competitive channels like the dead remnants of the first-star generation \citep{vanWassenhove10}, the~gravitational runaway of the densest stellar clusters \citep{Giersz2015}, or~the direct collapse of pristine gas within massive halos \citep{Volonteri12, Bonoli14}. They shed light on the connections with nuclear star clusters [NSCs];~\citep{Neumayer20}, the~behavior of BH--galaxy scaling relations at the low-mass regime \citep{Greene20, Inayoshi20}, the~origins of ultracompact dwarf galaxies as former galaxy nuclei \citep{Seth14, Voggel18}, and predictions of the gravitational-wave detection rates for the Laser Interferometer Space Antenna [LISA];~\citep{Amaro-Seoane23}, which can detect longer-wavelength radiation than ground-based ones \citep{ArcaSedda21}. 

Additional support for massive BH seeds was the absence of direct observations of IMBHs with masses between those of stellar mass BHs and SMBHs \citep{Valluri05, Tremou18}, leading to the speculation that IMBHs might not exist. However, this rarity is now considered a potential sample selection bias \citep{Greene20, Neumayer20}. The~sphere of gravitational influence (SOI, $R_{\rm SOI} = GM_{\rm BH}/\sigma_\star^2$, where $G$ is the gravitational constant and $\sigma_\star$ is the central velocity dispersion of stars) around such IMBHs is typically too small to spatially resolve in most galaxies due to their further distance and small mass. For~example, a~IMBH with $M_{\rm BH} \approx 10^5\,M_\odot$ and $\sigma_\star \approx 30~\kms$ within $3.5$~Mpc has $R_{\rm SOI} \approx 0\farcs03$, which is smaller than the diffraction limits of current state-of-the-art 8--10 m class telescopes equipped with adaptive optics (AO) techniques, such as the Very Large Telescope (VLT) and Gemini (with a full width at half maximum (FWHM) of the point spread function (PSF), $\rm FWHM_{PSF} \approx 0\farcs1$--$0\farcs2$), or~Keck {($\approx$$ 0\farcs05$).} 

However, there is now a growing number of IMBH candidates at the centers of dwarf galaxies, based on indirect estimates of their masses \citep{Reines13, Baldassare15, Zaw20, Davis21}. \citet{Chilingarian18} found 305 distant IMBH candidates with $M_{\rm BH} =$ (0.3--2)$\times10^5$ M$_\odot$, 10 of which were confirmed as active galactic nuclei (AGN) via coincident X-ray emission. Furthermore, a~few IMBHs were directly detected in nearby dwarfs with bright NSCs \citep{Barth04, Thornton08, denBrok15, Nguyen14, Nguyen17, Nguyen18, Nguyen19, Nguyen22, Nguyen17conf, Davis20} or in globular clusters [GCs];~\citep{Noyola08, Noyola10, Feldmeier13, Kiziltan2017, Pechetti22, Haberle24}, although~these detections are often at the limit of what 8--10 m telescopes can~achieve. 

These observations are necessary for the exploration of the relationships between $M_{\rm BH}$ and various galaxy properties, including the stellar mass \citep{Magorrian98}, the~luminosity \citep{Kormendy95}, and~the stellar velocity dispersion or central velocity dispersion \citep{Ferrarese00} of the central spherical bulge of stars. These properties, once measured in other galaxies, can serve as indicators of $M_{\rm BH}$. Therefore, firm detections and measurements of these currently elusive IMBHs in the local universe are crucial for robustly constraining the BH--galaxy scaling correlations across the mass scale. These strongly suggest a close connection between the growth of SMBHs and the formation and evolution of their host galaxies \citep{Krajnovic18a}.

Given the importance of IMBHs and the current active races in searching for their existence, like the Measuring Black Holes in Below Milky Way mass galaxies Project [the MBHBM$_\star$];~\citep{Nguyen20, Nguyen21, Nguyen22, Ngo2025a}, {N25} presented a systematic feasibility study of dynamically detecting and measuring IMBHs in NSCs of nearby low-mass galaxies using realistic ELT/HARMONI simulations. They defined a well-motivated sample of 44 nucleated galaxies within 10~Mpc and constructed mock HARMONI observations for two representative systems spanning the distance and structural extremes of the sample. The~simulations extrapolated  Hubble Space Telescope (HST)/Wide Field Planetary Camera 2 (WFPC2) Planetary Camera (PC) surface-brightness (SB) profiles, resolved at $\sim$$0\farcs05$, down to the $\sim$$10$ milli-arcsecond (mas) scales probed by HARMONI by adopting a fixed inner core-Sérsic slope for the NSCs ($\gamma = 0.1$). Given that the intrinsic NSC light profiles on sub-HST scales remain observationally unconstrained, this assumption may not be universal and could introduce systematic uncertainties in the inferred IMBH masses, underscoring the importance of ELT-class imaging for directly resolving NSC structure on these~scales.

This framework was subsequently extended to larger distances, reaching the Virgo [$D \approx 16.5$~Mpc];~\citep{Ngo2025b} and Fornax [$D \approx 20$~Mpc]; \citep{Ngo2025c}{,} 
 clusters. These studies employed HST-based SB models, synthetic stellar population spectra, Jeans Anisotropic Modeling [JAM\endnote{{\tt jampy} {v7.2.0,} 
 accessed on 1 October 2025. \url{https://pypi.org/project/jampy/}.}] 
\citep{Cappellari08, Cappellari20}, and~end-to-end HARMONI simulations [HSIM\endnote{{v3.11,} 
 accessed on 1 October 2025.~\url{https://github.com/HARMONI-ELT/HSIM}.}] \citep{Zieleniewski15} to generate high-resolution near-infrared data cubes. They analyzed the mock data cubes as real data, using JAM to infer the IMBH mass and its uncertainty within a Bayesian framework, and~they demonstrated that, at~10~mas spatial resolution and feasible exposure times, HARMONI can robustly recover IMBH masses as low as $3\times10^{3}$--$3\times10^{4}$~M$_\odot$ within $D \lesssim 10$~Mpc, corresponding to resolving black holes with masses of order $\sim$0.5\% of the NSC mass. At~larger distances, detectability rapidly degrades, requiring IMBH masses $\gtrsim 10^{5}$~M$_\odot$ at $D \approx 20$~Mpc. Nevertheless, these results establish ELT/HARMONI as a transformative facility for constraining IMBH demographics in the nearby universe and for testing models of BH seed formation and low-mass BH--galaxy~coevolution.

In this work, we improve the stellar mass model for galaxies using SimCADO\endnote{{Available from} 
 accessed on 1 October 2025. \url{https://github.com/astronomyk/SimCADO}.} to simulate mock observations from the Multi-AO Imaging Camera for Deep Observations (MICADO) imager \citep{Davies10, Davies21} instead of extrapolating a 1D HST/WFPC2 surface-brightness profile. We will then use these new two-dimensional (2D) high-spatial-resolution images, combined with the mock HARMONI kinematics from {N25}, to~re-estimate the mass of IMBHs at the centers of nearby NSCs. This approach not only re-examines the feasibility of discovering IMBH populations for the upcoming large HARMONI IFS and MICADO imaging surveys, but~also validates accurate IMBH measurements by exploring the effect of variations in the inner surface-brightness profile of~NSCs.

We outline the use of SimCADO and describe how we employed it to model the $I$-band images expected to be observed with the MICADO imager for NGC~300 and NGC 3115 dw01, varying the inner power-law slope ($\gamma$) of the core-S\'ersic function used to represent the SB of these NSCs in Section~\ref{stellar_mass_model}. These two galaxies span the extremes of the distance range of the parent sample and were chosen to represent the overall characteristics and observational limits of the {N25} proposed HARMONI IMBH survey. Following this, we detail the mock kinematics measurements of these nuclei obtained from mock HARMONI IFS using HSIM from {N25}. We then explain the JAM, which we used to explore the effect of varying the inner SB of these NSCs on the IMBH mass measurements in Section~\ref{dynamical_model}. Finally, we discuss and summarize our findings in Section~\ref{Conclusion}.

\section{New Stellar Mass~Model}\label{stellar_mass_model}
\unskip

\subsection{MICADO~Imager}\label{micado}

MICADO is a first-generation ELT imager, designed to produce diffraction-limited imaging and long-slit spectroscopy at near-infrared (NIR) wavelengths. It operates with both a multi-conjugate laser guide star adaptive optics (MCAO) and a single-conjugate natural guide star adaptive optics (SCAO). MICADO can achieve a sensitivity comparable to that of the \textit{James Webb Space Telescope (JWST)} and offers a resolution improved by a factor of six. The~instrument can produce images covering a field-of-view (FoV) of {\mbox{$50.5\times 50.5~\mathrm{arcsec}^2$}} 
 at 0.82--2.45~\micron\ in wide-field mode, using an array of nine $4096\times 4096$-pixel detectors with a plate scale of $4~\mathrm{mas}$. In~zoom mode, with~a FoV of $18\times 18~\mathrm{arcsec}^2$, MICADO provides a plate scale of $1.5~\mathrm{mas}$ in high-contrast imaging mode. The~instrument can perform astrometry with $50~\mathrm{mas}$ precision for brighter sources and enables coronagraphy using focal and pupil plane coronagraphs. Additionally, its observation modes include a long-slit spectrograph with a spectral resolution of $\lambda/\Delta\lambda \approx 2.0\times 10^{4}$, covering two spectral ranges of 0.82--1.55~\micron\ and 1.49--2.45~\micron.

MICADO can address a broad range of modern astrophysics. Its science drivers focus on themes such as galaxy evolution through observations of high-redshift galaxies and relic populations in local galaxies, the~formation and evolution of galaxies in the early universe, and~the dynamics of dense stellar systems. Additionally, it aims to explore the full mass spectrum of BHs in extragalactic galaxies and at the center of the Milky Way, including SMBHs in galaxy nuclei and IMBHs in dense stellar clusters (e.g., NSCs and GCs). Other key areas include studying the star formation history of galaxies through resolved stellar populations, characterizing exoplanets, observing planet formation and circumnuclear disks at small angular scales, and~investigating various aspects of the Solar~System.

\subsection{SimCADO~Package}\label{simcado}

SimCADO is a data simulation package \citep{Davies16} designed to generate realistic mock detector plane array read-out files for MICADO. This {\tt Python 3} routine models the incoming light's optical path, representing elements in the optical train by taking into account the effects of the atmosphere, telescope, instrument, and~detector \citep{Leschinski16}. It can simulate the MICADO imaging with 4 mas and 1.5 mas per pixel in the wavelength range of 0.7--2.5~\micron\ and provide the users with raw data sets that will be similar to what MICADO will produce during a typical observing~run. 

SimCADO is also highly configurable in simulating various observational scenarios equipped with SCAO and MCAO and integrated with subsystem effects along the optical train (the performance of the derotator or atmospheric dispersion corrector). The~former effects are associated with the spectral dimension (1D in wavelength, $\lambda$) and the spatial dimension (2D in position, $(x, y)$). The~latter include shifts, rotations, convolutions, and~distortions, as~well as cross terms that couple the spatial and spectral components of the PSF for the incoming photons. Additionally, SimCADO additionally includes several realistic features for simulating mock imagery for the MICADO detector array like the world coordinate system, variable sky background, PSF variability over the FoV, instrumental distortion map, missing segments due to the mirror surface's re-coat, and~extra-terrestrial optical path elements because of Zodiacal light, galactic extinction, atmospheric extinction, and~scattered~moonlight.

In practice, SimCADO accepts inputs in FITS or ASCII formats, and~it outputs all data as standard FITS files, allowing flexibility with other programs already in use within the astronomical~community.

\subsection{Inner Surface-Brightness Profile~Variation} \label{variation_mass_model}

\subsubsection{Mock MICADO Image~Simulations}\label{simcado_sim}

The range of inner slopes $\gamma$ explored in this work is motivated by theoretical considerations of the stellar dynamical equilibrium \citep{Tremaine94} and recent simulations by {N25}, rather than by direct observational constraints, which are currently unavailable on the relevant spatial scales. These studies show that stellar systems with $\gamma>0$ and no central black hole exhibit a projected velocity-dispersion profile that decreases toward the nucleus, whereas the presence of a black hole produces a central rise in $\sigma_p$ within the sphere of influence, asymptotically approaching the Keplerian form $\sigma_p \propto R^{-1/2}$ for $0<\gamma<2$, largely independent of the underlying density slope. Physically plausible stellar cusps therefore span $0<\gamma<2$. While {N25} adopted a fiducial value of $\gamma = 0.1$, reflecting the expectation that realistic NSCs are unlikely to have perfectly flat cores or extremely steep cusps owing to mass segregation and core contraction effects—particularly when probed by ELT-class facilities—we here explore a broader range, $0 \leq \gamma \leq 0.7$, to~quantify how increasing central stellar density affects the robustness and limitations of stellar-dynamical IMBH mass measurements with ELT/MICADO and~HARMONI.

We used SimCADO to simulate the $I$-band images for NGC~300 and NGC~3115~dw01 observed with MICADO in the wide-field mode. Here, we used the best-fit parameters of their NSCs' SB constrained from the HST/WFPC2 F814W images as discussed in {Section~5} of {N25} 
 and summarized in {Table~4} of {N25}. 

\begin{figure}[H]

\begin{adjustwidth}{-\extralength}{0cm}
\centering 
\includegraphics[width=1.1\textwidth]{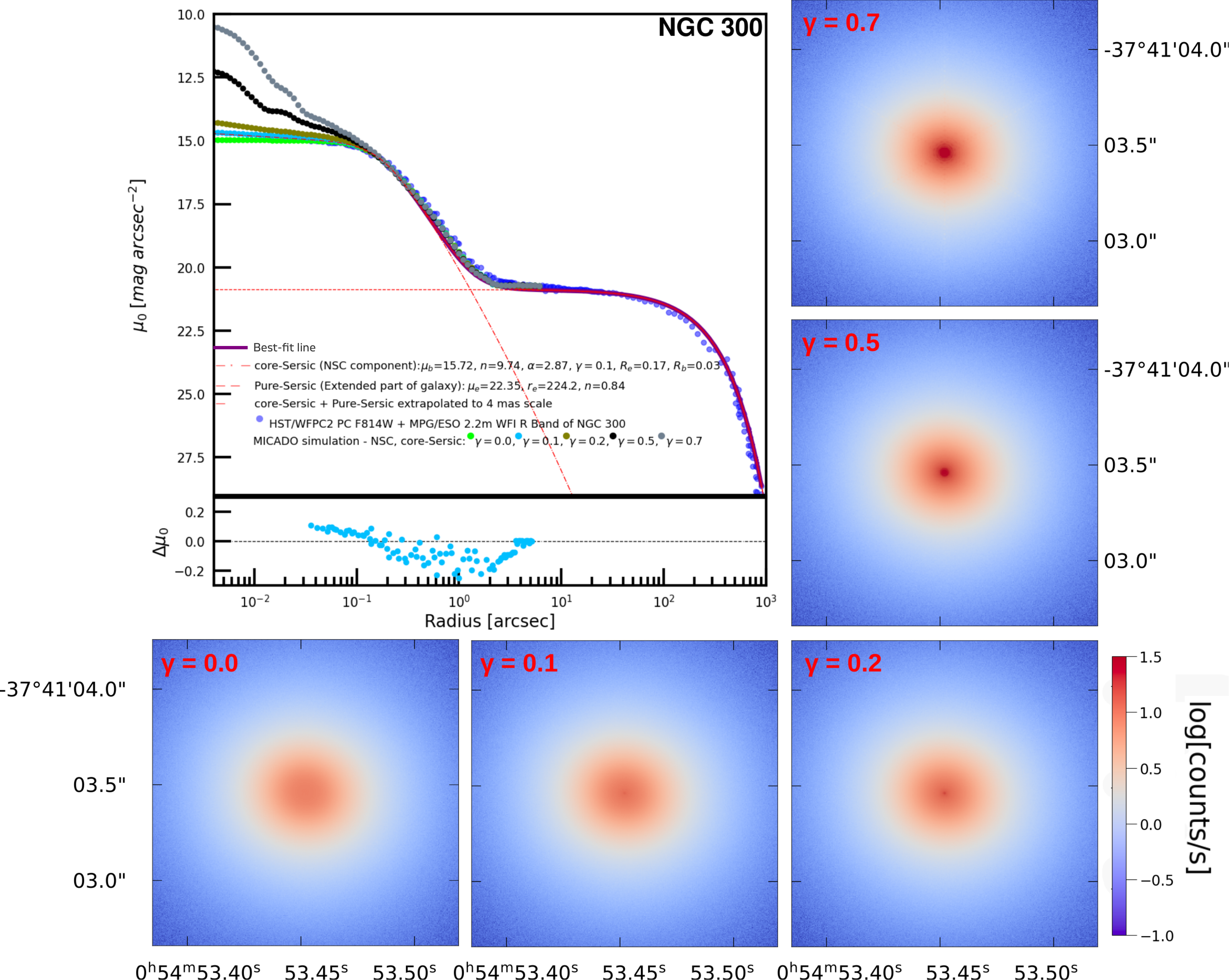}
\end{adjustwidth}
    \caption{{\it {1D} 
 profiles:} of the  HST/WFPC2 F814W and MPG/ESO 2.2-m $R$ SB of NGC 300, constructed directly from IRAF 
 ellipse (blue dots), taken from {N25}. The~best-fit core-S\'ersic + S\'ersic SB extrapolated to the 4 mas scale are plotted as thick purple solid lines, with~their best-fitting parameters shown in the legend. The~core-S\'ersic profile is represented by a red dotted line, and~the S\'ersic profile is shown by a red dashed line. We overlaid the MICADO $I$-band images with different inner-power law slopes $\gamma=$ 0 (green), 0.1 (cyan), 0.2 (olive), 0.5, (black), and~0.7 (gray). {\it Residual profile:} We display the differences ({\tt data $-$ 
 model}) between the {\tt IRAF ellipse} SB and the corresponding MICADO $I$-band images with the inner-power law slope of $\gamma=0.1$ in the associated colors to demonstrate their radial agreements/disagreements. {\it $I$-band MICADO images:}  The MICADO $I$-band images with different inner-power law slopes in the same color-bar~scale.}
    \label{mass-model_NGC300}
\end{figure}

\begin{figure}[H]

\begin{adjustwidth}{-\extralength}{0cm}
\centering 
\includegraphics[width=1.1\textwidth]{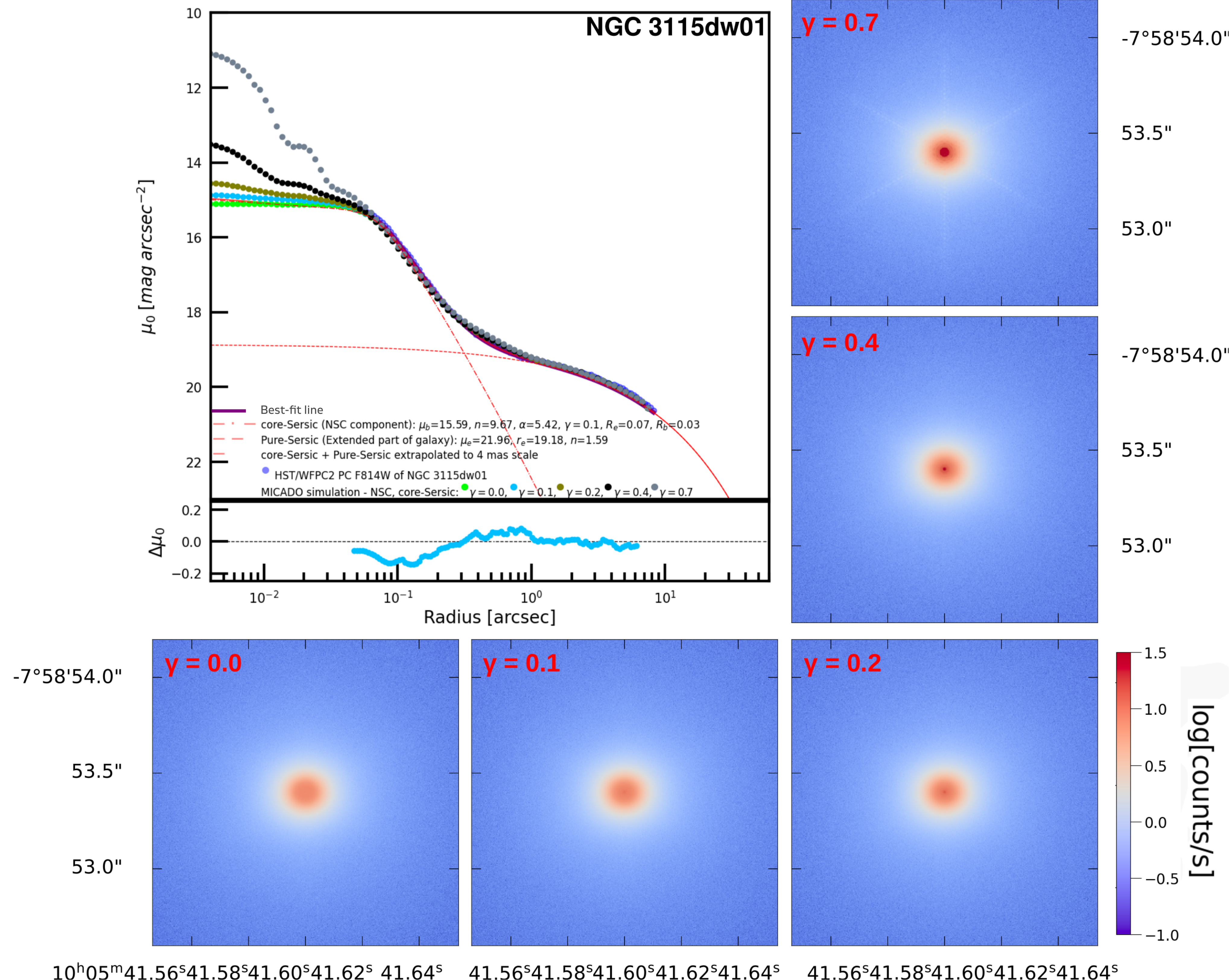}
\end{adjustwidth}
    \caption{{Same} 
 as Figure~\ref{mass-model_NGC300} with the HST/WFPC2 F814W SB constructed directly from IRAF ellipse for NGC~3115~dw01 taken from {N25}. The~figure also shows the MICADO $I$-band images and their 1D SB with different inner-power law slopes in the same color-bar~scale.}
    \label{mass-model_NGC3115dw01}  
\end{figure}

{N25} adopted a combination of a core-S\'ersic and a S\'ersic model to fit the NSC and the extended component of these galaxies, respectively. The~core-S\'ersic model \citep{Graham03, Trujillo04} parametrizes the projected SB of NSCs, rather than their intrinsic density, {as~follows:} 
\begin{equation}
	I(r) = I' \left[1 + \left(\frac{R_b}{r}\right)^\alpha\right]^{\gamma/\alpha} 
	\exp\left[{-b_{n_{\rm 1}} \left(\frac{r^\alpha + R_b^\alpha}{R_e^\alpha}\right)^{1/(\alpha\, n_{\rm 1})}}\right] 
\end{equation} 
where 
\begin{equation*}
I'=I_b 2^{-\gamma/\alpha}\exp\left[b_{n_{\rm 1}}2^{1/(\alpha\, n_{\rm 1})}\left( \dfrac{R_b}{R_e}\right)^{1/n_{\rm 1}} \right]
\end{equation*} 
{In} 
 this function, the~S\'ersic index ($n_{\rm 1}$) and the power-law slope ($\gamma$) control the shapes of the outer S\'ersic part and the inner power-law regime, while the sharpness parameter ($\alpha$) determines the transition between these two regions. $R_b$ is the break radius at which the transition occurs, and~thus $I_b$ is the intensity at $R_b$ (converted to SB $\mu_b$ in the legends of Figures~\ref{mass-model_NGC300} and \ref{mass-model_NGC3115dw01}). Outside $R_b$, the~profile follows a \citet{Sersic68} profile with a projected half-light radius $R_e$, but~it gradually transitions to a power-law SB $I(r) \propto r^{-\gamma}$ at smaller radii $r \ll R_b$. Finally, the~parameter $b_{n_{\rm 1}}$ is given {as} 
  $b_{n_{\rm 1}} \approx 2n_{\rm 1} - \dfrac{1}{3} + \dfrac{4}{405n_{\rm 1}} + \dfrac{46}{25515n^2_{\rm 1}} + {\rm O}\left(\dfrac{1}{n^3_{\rm 1}}\right)$~\citep{Ciotti99}.

On the other hand, the~\citet{Sersic68} model used to describe the extended disk of these galaxies has a form:
\begin{equation} 
	I(r) = I_e\exp \Biggl\{-b_{n_{\rm 2}}\left[{\left(\frac{r}{r_e}\right)^{1/n_2}}-1\right]\Biggr\}	 
\end{equation} 
where $n_2$ represents the S\'ersic index, distinguishing it from $n_{\rm 1}$ of the core-S\'ersic profile in Equation~(1), and~it relates to the coefficient $b_{n_{\rm 2}} \approx 2n_{\rm 2} - 1/3$. Here, $r_e$ is the effective radius of the profile, and~$I_e$ is the intensity at $r_e$, which is converted to SB $\mu_e$, as~shown in the legends of Figures~\ref{mass-model_NGC300} and \ref{mass-model_NGC3115dw01}.

The cusp in the inner slope ($\gamma$) of the core-S\'ersic profile reflects our current lack of knowledge about the profiles of NSCs at small radii. These regions are inaccessible with current facilities but will become observable with the ELT. {N25} adopted a power-law index of $\gamma=0.1$ in their extrapolated SB and simulations at these radii. In~this work, we modeled the $I$-band MICADO images based on the best-fit parameters of these galaxies' SB, but~we allowed $\gamma$ to vary: $\gamma=0,\ 0.1,\ 0.2,\ 0.5,\ 0.7$ for NGC~300 and $\gamma=0,\ 0.1,\ 0.2,\ 0.4,\ 0.7$ for NGC~3115~dw01. 

We clarify the choice of $\gamma$ for both galaxies. The~simulations with $\gamma=0$ serve to validate predictions where the inner SB approaches a constant value, and~the velocity dispersion profile asymptotically becomes a constant positive value, rather than declining toward zero—even in the absence of an IMBH, as~demonstrated in earlier numerical \citep{Tremaine94} and simulation work ({N25}). Simulations with $\gamma=0.1$ serve to assess the consistency of the galaxy mass models generated from the mock MICADO $I$-band images in this study with those interpolated into the 4 mas regime, derived from the HST/WFPC2 F814W images. For~$\gamma>0.1$, we explore how $\gamma$ variation affects the determination of IMBH mass through stellar dynamics when using MICADO~data.

We enhanced the SimCADO package by adding a 2D core-S\'ersic function (Equation~(1)), complementing the existing 2D S\'ersic function (Equation~(2)). During~our simulations, we first generated emission curves based on the internal spectral energy distribution (SED) for an elliptical galaxy (NGC~3115~dw01) and a spiral galaxy (NGC 300) in the $I$-band filter. We then adjusted the spectral dimension to longer wavelengths, corresponding to the redshifts of NGC~300 ($z = 0.00048$) and NGC~3115~dw01 ($z = 0.00227$), as~reported on the NASA/IPAC Extragalactic Database (NED\endnote{{ \url{https://ned.ipac.caltech.edu}, accessed on 1 October 2025.}}). 
 Although~these redshifts are negligible, this step was included for~accuracy. 

For each galaxy, we created 2D models using the core-S\'ersic profile for the NSC ($q=0.999$) and the S\'ersic profile for the extended-disk component ($q=0.999$ for NGC~300 and elliptical with $q=0.9$ for NGC~3115~dw01), and we then combined these models into a single source. We also fixed certain observational parameters to mimic the realistic observations, including {\tt obs\_dit = 900~s} (observing time per exposure) and {\tt n\_dit = 4} (number of~exposures). 

\subsubsection{PSF Determination from MICADO Point Source~Images}\label{sec:psf_determination_from_simulated_micado_point_source_images}

Accurate characterization of the PSF is essential for deconvolving galaxy light distributions and performing reliable photometric and structural analyses. For~MICADO, the~PSF is expected to be complex and sensitive to both AO performance and observing~conditions.

To model the PSF, we simulate MICADO observations of point sources (i.e., stars) using SimCADO, adopting the same instrumental setup and conditions as those used for the galaxy simulations. The~resulting stellar images are analyzed with the Multi-Gaussian Expansion (MGE) method, implemented through the {\tt mge.fit\_sectors} routine from the \textsc{MgeFit} {Python} 
 package\endnote{{v5.0:} 
 accessed on 1 October 2025. \url{https://pypi.org/project/mgefit/}.} \citep{Cappellari02}. The~{\tt mge.fit\_sectors} algorithm models the 2D light distribution of each simulated star as a sum of co-axial Gaussian components. This simulated-MGE PSF is subsequently incorporated into the analysis of mock MICADO galaxy images to accurately account for instrumental and atmospheric broadening~effects.

\subsection{Brightness Profiles from MICADO Galaxy~Images}\label{sec:mge_surface_brightness_parametrization_from_simulated_micado_galaxy_images}

We showed in Figure~\ref{mass-model_NGC300} for NGC~300 and Figure~\ref{mass-model_NGC3115dw01} for NGC~3115~dw01 the $I$-band MICADO images for varying the inner power-law slope  of the core-S\'ersic, as~well as their 1D SB extracted from the {\tt Image Reduction and Analysis Facility (IRAF) ellipse} task~\citep{Jedrzejewski87}. The~{\tt ellipse} routine systematically integrated the flux of stars within concentric annuli, allowing for variations in position angles and ellipticities along the galaxy's semi-major axis and deconvolving with the simulated PSF image (Section~\ref{sec:psf_determination_from_simulated_micado_point_source_images}). Subsequently, we converted the average flux within each annulus (measured in counts/s) into SB expressed in mag arcsec$^{-2}$. The~photometric zero point for the $I$ band is {\tt \mbox{zero\_point$_I$ = 29.492}} mag obtained from the photometric zero point function within the SimCADO framework, which simulated the light from a point source (i.e., a~star) and \mbox{calculated} as {\tt zero\_point$_I$ = 2.5$\log_{10}$(counts/t\_exp) + input\_mag}; here, {\tt t\_exp = 1s} and {\tt input\_mag = 10} mag. 

To highlight the effects of varying $\gamma$ of the core-Sérsic profile on the SB at radii probed only by the ELT, we overlaid the HST/WFPC2 F184W {\tt IRAF ellipse} profiles and their best-fit core-Sérsic + Sérsic functions from {N25} in the same figure for each galaxy. It is evident that at these unexplored radii, larger $\gamma$ values significantly increase the light distribution of stars within the NSC towards its center. This change will indeed impact the determination of the IMBH mass (Section~\ref{IMBHmass_measurement}). To~validate our simulations, we compared the previous profile with $\gamma=0.1$ fitted by {N25} and found these profiles consistent to each other in both galaxies. This consistency is also evident in the case of $\gamma=0$, where our simulations produced flat profiles toward the galaxy centers, as~demonstrated in previous numerical work by \citet{Tremaine94} and simulations ({Figure 4 and Section~4.2} 
 of {N25}).

\begin{figure}[H]
    \includegraphics[width=0.65\textwidth]{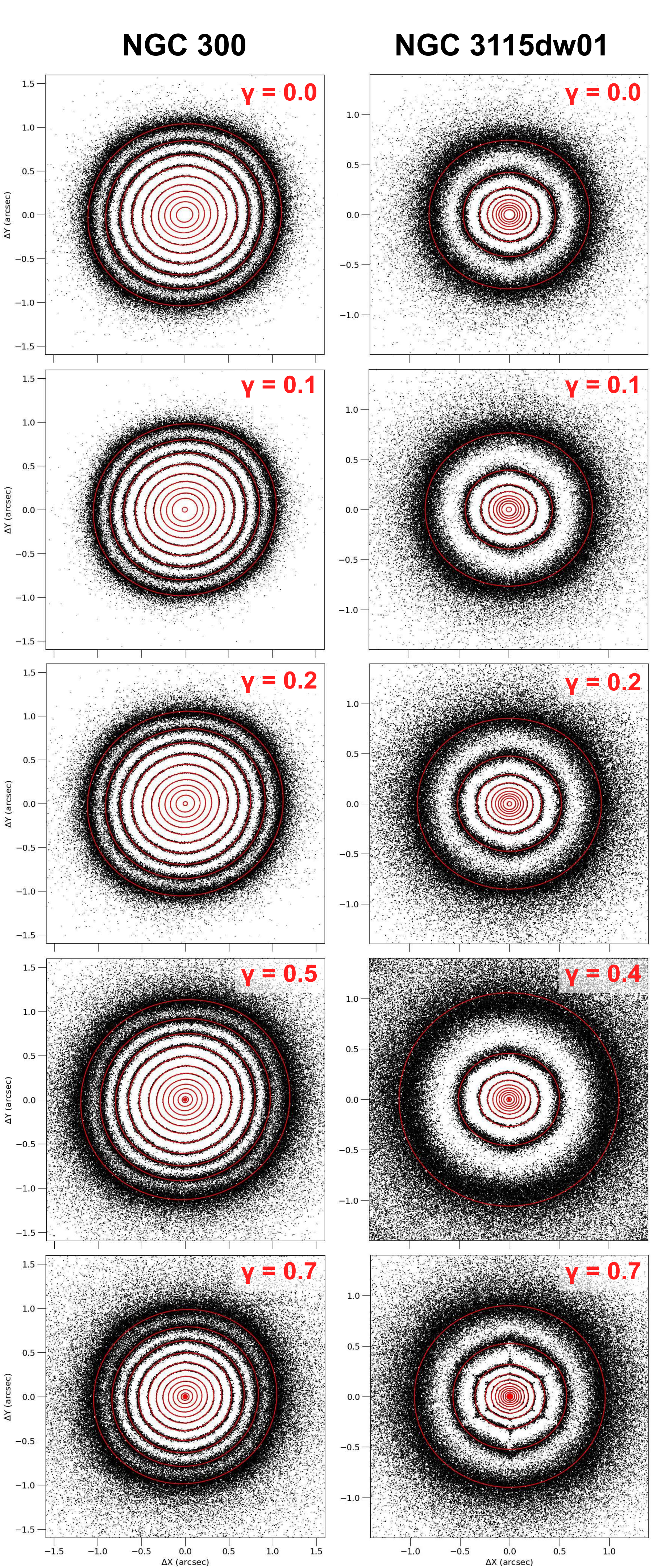}
    \caption{Comparisons between the MICADO $I$-band images produced by SimCADO and their best-fit MGE models for NGC~300 (\textbf{left}) and NGC~3115~dw01 (\textbf{right}) with various inner power-law indices is presented in terms of 2D SB density. Black data points represent the mock data, while red contours depict the MGE models, illustrating the alignment between data and model at corresponding radii and contour~levels.}
    \label{mges_sim}  
\end{figure}

We converted these MICADO $I$-band images to MGE models using the {\tt Python} version of the {\tt MgeFit} with the {\tt mge\_fit\_sectors\_regularized} routine. During~the fits, we also convolved the images with the MICADO $I$-band PSF. We saved these MGE-light models of the two galaxies varying with the inner power-law slope ($\gamma$) in Table~\ref{mgetab} and illustrated in Figure~\ref{mges_sim}. This Figure shows the agreement/disagreement between the data and the model as 2D contours at equivalent radii and contour levels for the mock MICADO simulations with five simulated $\gamma$ for each galaxy. These MGE models can be analytically de-projected with a specified axis ratio and inclination to reconstruct the three-dimensional (3D) distribution of the entire~galaxy. 

Finally, we converted the luminosity surface density into mass surface density by assuming a constant mass-to-light ratio (\ml). For~NGC 300, we used $M/L_{\rm F814W, dyn}\approx M/L_I\approx0.6$ (\Msun/\Lsun) as reported by \citet{Neumayer12}, and~for NGC 3115 dw01 we adopted $M/L_{\rm F814W, phot}\approx M/L_I\approx1.4$ (\Msun/\Lsun) derived by \citet{Pechetti20}.

\begin{table}[H]
\caption{{MGE} 
 models of NGC~300 and NGC~3115~dw01 modeled from SimCADO for mock MICADO $I$-band~images.}
\label{mgetab}

\begin{adjustwidth}{-\extralength}{0cm}

\setlength{\cellWidtha}{\fulllength/7-2\tabcolsep}
\setlength{\cellWidthb}{\fulllength/7-2\tabcolsep}
\setlength{\cellWidthc}{\fulllength/7-2\tabcolsep}
\setlength{\cellWidthd}{\fulllength/7-2\tabcolsep}
\setlength{\cellWidthe}{\fulllength/7-2\tabcolsep}
\setlength{\cellWidthf}{\fulllength/7-2\tabcolsep}
\setlength{\cellWidthg}{\fulllength/7-2\tabcolsep}

\begin{tabularx}{\fulllength}{
  >{\centering\arraybackslash}m{\cellWidtha}
  >{\centering\arraybackslash}m{\cellWidthb}
  >{\centering\arraybackslash}m{\cellWidthc}
  >{\centering\arraybackslash}m{\cellWidthd}
  >{\centering\arraybackslash}m{\cellWidthe}
  >{\centering\arraybackslash}m{\cellWidthf}
  >{\centering\arraybackslash}m{\cellWidthg}
  >{\centering\arraybackslash}m{\cellWidthd}
}
\toprule
  & \boldmath$\lg \Sigma_{\star,j}/(\Lsun\,{\rm pc^{-2}})$ 
  & \boldmath$\sigma_j$ ($\arcsec$) 
  & \boldmath$q'_j=b_j/a_i$
  & \boldmath$\lg \Sigma_{\star,j}/(\Lsun\,{\rm pc^{-2}})$ 
  & \boldmath$\sigma_j$ ($\arcsec$) 
  & \boldmath$q'_j=b_j/a_i$ \\
 \textbf{ (1)}   &   \textbf{(2)}  & \textbf{(3)} &  \textbf{(4)} & \textbf{(5)}  &  \textbf{(6)} & \textbf{(7)} \\	                
\midrule
\boldmath$j$  &    &  \textbf{NGC~300}   &  &   & \textbf{NGC~3115~dw01} & \\
\midrule
-- &     & $\gamma=0$ &      &   & $\gamma=0$ &  \\
\midrule 
1 & 4.344 & 0.122 & 0.999 & 4.343  & 0.134 & 0.999 \\ 
2 & 3.469 & 0.126 & 0.999 & 3.469  & 0.286 & 0.999 \\ 
3 & 1.902 & 0.238 & 0.999 & 1.902  & 0.719 & 0.999 \\ 
4 & 1.048 & 0.249 & 0.999 & 1.048  & 0.830 & 0.90 \\ 
5 & 1.072 & 0.562 & 0.999 & 1.072  & 1.626 & 0.90 \\ 
6 &  --   &   --  &  --   &$-$0.021& 2.318 & 0.90 \\
7 &  --   &   --  &  --   & 0.172  & 3.909 & 0.90 \\
8 &  --   &   --  &  --   & 0.213  & 6.944 & 0.90 \\
9 &  --   &   --  &  --   &$-$0.096&11.015 & 0.90 \\

\bottomrule
\end{tabularx}
\end{adjustwidth}
\end{table}

\begin{table}[H]\ContinuedFloat

\caption{{\em Cont.}}
\begin{adjustwidth}{-\extralength}{0cm}
\label{mgetab}

\setlength{\cellWidtha}{\fulllength/7-2\tabcolsep}
\setlength{\cellWidthb}{\fulllength/7-2\tabcolsep}
\setlength{\cellWidthc}{\fulllength/7-2\tabcolsep}
\setlength{\cellWidthd}{\fulllength/7-2\tabcolsep}
\setlength{\cellWidthe}{\fulllength/7-2\tabcolsep}
\setlength{\cellWidthf}{\fulllength/7-2\tabcolsep}
\setlength{\cellWidthg}{\fulllength/7-2\tabcolsep}

\begin{tabularx}{\fulllength}{
  >{\centering\arraybackslash}m{\cellWidtha}
  >{\centering\arraybackslash}m{\cellWidthb}
  >{\centering\arraybackslash}m{\cellWidthc}
  >{\centering\arraybackslash}m{\cellWidthd}
  >{\centering\arraybackslash}m{\cellWidthe}
  >{\centering\arraybackslash}m{\cellWidthf}
  >{\centering\arraybackslash}m{\cellWidthg}
  >{\centering\arraybackslash}m{\cellWidthd}
}
\toprule
  & \boldmath$\lg \Sigma_{\star,j}/(\Lsun\,{\rm pc^{-2}})$ 
  & \boldmath$\sigma_j$ ($\arcsec$) 
  & \boldmath$q'_j=b_j/a_i$
  & \boldmath$\lg \Sigma_{\star,j}/(\Lsun\,{\rm pc^{-2}})$ 
  & \boldmath$\sigma_j$ ($\arcsec$) 
  & \boldmath$q'_j=b_j/a_i$ \\
 \textbf{ (1)}   &   \textbf{(2)}  & \textbf{(3)} &  \textbf{(4)} & \textbf{(5)}  &  \textbf{(6)} & \textbf{(7)} \\	                
\midrule
\boldmath$j$  &    &  \textbf{NGC~300}   &  &   & \textbf{NGC~3115~dw01} & \\
\midrule
-- &    &$\gamma=0.1$&  &   &$\gamma=0.1$&\\
\midrule
1 & 3.648 & 0.005 & 0.999 & 4.177  & 0.006 & 0.999 \\ 
2 & 3.351 & 0.017 & 0.999 & 3.313  & 0.113 & 0.999 \\ 
3 & 2.841 & 0.027 & 0.999 & 2.911  & 0.108 & 0.999 \\ 
4 & 4.247 & 0.036 & 0.999 & 4.362  & 0.125 & 0.999 \\ 
5 & 4.081 & 0.088 & 0.999 & 2.958  & 0.273 & 0.999 \\ 
6 & 2.623 & 0.110 & 0.999 & 2.315  & 0.732 & 0.90 \\ 
7 & 3.565 & 0.177 & 0.999 & 1.048  & 0.830 & 0.90 \\ 
8 & 3.330 & 0.503 & 0.999 & 1.072  & 1.626 & 0.90 \\ 
9 &   --  &   --  &  --   &$-$0.021& 2.318 & 0.90 \\
10&   --  &   --  &  --   & 0.172  & 3.909 & 0.90 \\
11&   --  &   --  &  --   & 0.213  & 6.944 & 0.90 \\
12&   --  &   --  &  --   &$-$0.096&11.015 & 0.90 \\
\midrule
-- &   &$\gamma=0.2$&   &   &$\gamma=0.2$&\\
\midrule
1 & 4.218 & 0.005 & 0.999 & 3.954  & 0.004 & 0.999 \\ 
2 & 3.848 & 0.017 & 0.999 & 3.890  & 0.012 & 0.999 \\ 
3 & 3.443 & 0.035 & 0.999 & 4.374  & 0.111 & 0.999 \\ 
4 & 3.927 & 0.080 & 0.999 & 3.517  & 0.231 & 0.999 \\ 
5 & 3.921 & 0.101 & 0.999 & 2.315  & 0.732 & 0.90 \\ 
6 & 3.876 & 0.128 & 0.999 & 1.048  & 0.830 & 0.90 \\ 
7 & 2.829 & 0.135 & 0.999 & 1.072  & 1.626 & 0.90 \\ 
8 & 3.850 & 0.176 & 0.999 &$-$0.021& 2.318 & 0.90 \\ 
9 & 3.600 & 0.285 & 0.999 & 0.172  & 3.909 & 0.90 \\
10& 3.341 & 0.562 & 0.999 & 0.213  & 6.944 & 0.90 \\
11&  --   &  --   &  --   &$-$0.096&11.015 & 0.90 \\
\midrule
-- &    &$\gamma=0.5$&  &   &$\gamma=0.4$&\\
\midrule
1 & 5.574 & 0.005 & 0.999 & 4.823  & 0.005 & 0.999 \\ 
2 & 4.664 & 0.017 & 0.999 & 4.425  & 0.017 & 0.999 \\ 
3 & 4.116 & 0.046 & 0.999 & 4.281  & 0.073 & 0.999 \\ 
4 & 3.179 & 0.059 & 0.999 & 4.408  & 0.101 & 0.999 \\ 
5 & 4.193 & 0.089 & 0.999 & 2.456  & 0.103 & 0.999 \\ 
6 & 3.804 & 0.117 & 0.999 & 3.468  & 0.250 & 0.999 \\ 
7 & 3.993 & 0.167 & 0.999 & 2.315  & 0.732 & 0.90 \\
8 & 2.077 & 0.332 & 0.999 & 1.048  & 0.830 & 0.90 \\ 

\bottomrule
\end{tabularx}

\end{adjustwidth}

\noindent{\footnotesize{\textit{Notes:} MGE models with different inner power-law slopes of the core-S\'ersic function of NSCs. Each model has a specific number of Gaussian as shown in Column 1. Columns 2--4: The light-SB density, the~Gaussian width along the galaxy major axis, and~the ratio between the semi-minor and semi-major axes for NGC 300. Columns 5--7: Similarities of Columns 2--4 but for NGC~3115~dw01.} }
\end{table}
\unskip

\section{Dynamical~Model}\label{dynamical_model}
\unskip

\subsection{Jeans Anisotropic Model (JAM)}\label{jam}

We utilize the JAM model based on the solution of the Jeans equations, assuming axisymmetry with a cylindrical alignment of the velocity ellipsoid, which is axially symmetric around the vertical direction ($\sigma_z \neq \sigma_r = \sigma_\phi$). This JAM-predicted LOSVD is calculated in combination with the new mass model simulated by SimCADO (Section~\ref{stellar_mass_model}) before being compared to the mock kinematics generated by {N25} through a Bayesian framework to infer the IMBH mass and its~uncertainty. 

This choice of cylindrical aligned JAM is to ensure consistency with the mock kinematics generated by {N25}, where they used the same model for their HARMONI IMBH sample of dwarf galaxies hosting bright NSCs with significant rotational characteristics. We implemented this by setting {\tt align=‘cyl’} in the \textsc{jam\_axi\_proj} procedure of the {\textsc{JamPy}} 
 {package} 
 \citep{Cappellari20}.

\subsection{Mock Stellar~Kinematics}\label{stellar_kinematics}

This work used the stellar kinematics measured from the mock data cubes produced from HSIM. The~detailed descriptions of how these mock observations were produced in various spectral gratings of medium ($\lambda/\Delta\lambda\approx7100$; $J$, $H$, $K$) and high ($\lambda/\Delta\lambda\approx$~17,400; $H$-high, $K$-short, $K$-long) spectral resolution have presented in {Sections~6.2 and 6.3} 
 of {N25}. In~brief, {N25} used HSIM to generate mock HARMONI datacubes for NGC~300 and NGC~3115~dw01 at the 10~mas spaxel scale, assuming three IMBH masses (\Mbh~=~0\%, 0.5\%, and~1\%\Mnsc) and extracted the stellar kinematics ($V$, $\sigma_\star$) from these datacubes using \textsc{pPXF} across all six~gratings.

{N25} showed these stellar kinematic measurements from their mock HARMONI datacubes simulated for three different \Mbh: =~0\%\Mnsc, =0.5\%\Mnsc, and~=1\%\Mnsc\ for NGC~300 and NGC 3115 dw01, respectively. The~adoption of fixed fractional black hole masses relative to the host NSC in the mock HARMONI simulations reflects the absence of a well-established \Mbh–galaxy scaling relation in the low-mass regime that would allow reliable predictions of \Mbh\ from host galaxy properties. {N25} therefore parameterized the black hole mass as a fixed fraction of the NSC mass, \Mnsc. Given that the proposed HARMONI IMBH survey is restricted to distances $\lesssim 10$~Mpc, the~lower bound of this fractional mass range was chosen to ensure that the IMBH sphere of influence remains resolvable at the nominal 10~mas spatial resolution. This criterion guarantees that the stellar kinematics in the central spaxels are dominated by the gravitational potential of the IMBH rather than that of the surrounding NSC. The~upper bound was selected to bracket a plausible range of IMBH masses consistent with existing observational~constraints.

Furthermore, the~two adopted fixed values of $M_{\rm BH}/M_{\rm NSC}=$ 0.5\% and 1\% were chosen to probe the minimum IMBH masses that can be robustly detected with HARMONI stellar-dynamical observations, given the N25 survey distances (2~Mpc for NGC~300 and 9.7~Mpc for NGC~3115~dw1) and the nominal 10~mas spatial resolution. When the black hole sphere of influence is marginally resolved, the~recovered IMBH masses show no significant systematic bias and are consistent with the input values within the uncertainties. At~lower mass fractions, where the sphere of influence is unresolved, the~inferred IMBH masses become increasingly biased and converge toward effective upper limits rather than well-constrained detections, reflecting the loss of sensitivity of the stellar-dynamical method in this~regime.

These stellar kinematic maps (i.e., with~an assumed $\gamma = 0.1$) show distinct properties of the presence or absence of an IMBH, exhibiting distinctive central drops in $\sigma_\star$ and $V_{\rm rms}$ (if nuclei are less rotating) maps when either no BH ($M_{\rm BH}=0$ \Msun) is present or when the $M_{\rm BH}$ are below a certain threshold (e.g., $M_{\rm BH}=5\times10^3$ \Msun\ for NGC~300 and $M_{\rm BH}=3.5\times10^4$ \Msun\ for NGC 3115 dw01; or \Mbh~=~0.5\%\Mnsc). This central-$\sigma_\star$ drop is a common feature in the predicted stellar kinematics of galaxies when no or very small IMBHs are assumed, and~it holds true for a range of assumed anisotropic profiles and density profiles with $\gamma > 0$ regardless of the Sérsic index (\citet{Tremaine94}; {N25}). This well-established feature of realistic galaxy models initially appeared to contradict early observations \citep{Binney80}. However, the~apparent discrepancy was resolved when it was discovered that all massive galaxies contain SMBHs. Conversely, models featuring relatively large IMBHs generate centrally rising peaks in the $\sigma_\star$ map at several central spaxels in the Keplerian manner $\sigma_\star\propto r^{-1/2}$ (for $0<\gamma<2$) irrespective of the profile slope and Sérsic index, which correspond to the BH’s SOI, showing a clearly upturned trending as $M_{\rm BH}$ increases.   This observation aligns with the general expectation that central $\sigma_\star$ should increase in a Keplerian manner within a compact tiny region near the central IMBH, where its gravitational potential dominates over all~others.

Given the highly consistent mock stellar kinematics presented by {N25} across six spectral gratings and two spectral resolutions, we adopt the stellar kinematics produced from the $H$-high grating to test the variation effect of the inner SB of NSCs simulated with MICADO on IMBH mass measurements~only.

\subsection{Improve IMBH Mass Measurements with~MICADO}\label{IMBHmass_measurement}

We performed fitting of the JAM to mock $H$-high kinematics ($V_{\rm rms}$), aimed at testing the variation effect of the inner-power law's slope ($\gamma$) of the core-S\'ersic function of NSCs simulated with MICADO on IMBH mass determinations and at constraining the model’s dynamical parameters.  Here, we replaced the interpolated HST mass model toward the central 4-mas region produced by {N25} by the 2D MICADO images simulated from the SimCADO with varying a range value of $\gamma$ (Section~\ref{variation_mass_model}). 

The JAM is characterized with a central IMBH  mass ($M_{\rm BH}$), the~stellar orbitals represented by the anisotropy ($\beta_z$), the~galaxy morphological inclination angle ($i$), and~the constant stellar mass-scaling factor (\ml$_{I}$). We adopted a logarithmic scale for $M_{\rm BH}$ to ensure a more Gaussian sampling across multiple orders of magnitude, while the other three parameters are sampled in linear scales. During~the fitting process, JAM generated kinematic models that could be compared to their corresponding simulated values ($V_{\rm rms}$) while accounting for the HARMONI LTAO PSF with a FWHM$_{\rm PSF}\approx12$ mas. 

To obtain the best-fit JAM, we utilized a Markov chain Monte Carlo (MCMC) simulation to explore the parameter space of $M_{\rm BH}$, \ml$_{I}$, $\beta_z$, and~$i$, fitting the mock kinematic data and determining the best-fit parameters and uncertainties (statistical and kinematic measurement), using the adaptive Metropolis algorithm \citep{Haario01} within the Bayesian framework [{\tt adamet}\endnote{v2.0.9 {available from} 
 accessed on 1 October 2025. \url{https://pypi.org/project/adamet/}.}];~\citep{Cappellari13a}. Our MCMC chains comprised of $3\times10^4$ iterations, with~the initial 20\% of iterations excluded as a burn-in phase. The~full probability distribution function (PDF) was constructed from the remaining 80\% of calculations. The~best-fit parameters are the highest likelihood of the PDF, while the uncertainties for all four parameters were determined at 1$\sigma$ and 3$\sigma$ confidence levels, representing the ranges (16–84)\% and (0.14–99.86)\% of the PDF, respectively. We fixed the search ranges for four model parameters as follows:
\begin{itemize}
\item$\lg(M_{\rm BH}/$\Msun): from 0 to 6 (or $M_{\rm BH}$: from 0 to $10^6$ \Msun)
\item \ml$_{I}$: from 0.1 to 3 (\Msun/\Lsun)
\item $\beta_z$: from $-$1.0 to~0.99
\item$i$: from 5$^{\circ}$ to 90$^{\circ}$
\end{itemize}
Initial guesses for the JAM parameters were fixed with ( \ml$_{I}$, $\beta_z$, $i$) = (0.6, 0, 42) for NGC 300, while those = (1.4, 0, 42) for NGC 3115 dw01. The~ initial guesses for $M_{\rm BH}$ are those used in the simulated data cubes with $M_{\rm BH}=0,\;5\times10^{3},\;10^{4}$ \Msun\ for NGC~300 and $M_{\rm BH}=0,\;3.5\times10^{4},\;7\times10^{4}$ \Msun\ for NGC~3115~dw01. 

In Figures~\ref{new_ngc300_gamma_0.0}--\ref{new_ngc300_gamma_0.7} for NGC~300 and Figures~\ref{new3_ngc3115dw01_gamma_0.0}--\ref{new3_ngc3115dw01_gamma_0.7} for NGC~3115d~w01, the~best-fit JAM parameters and their associated uncertainties are shown. These parameters describe the derived stellar kinematics for three different $M_{\rm BH}$ and five inner power-law slopes $\gamma$. The~figures feature 2D scatter plots for each parameter, with~colored points indicating their likelihood. White represents the maximum likelihood, while black signifies a confidence level smaller than 3$\sigma$. Additionally, histograms are provided to display the 1D distributions for each parameter. These 1D distributions were used to determine the best-fit values and their corresponding uncertainties, which are listed in Table~\ref{jam_ngc300} for NGC~300 and Table~\ref{jam_ngc3115dw01} for NGC~3115~dw01. The~insert plots at the top-right corner of each PDF directly compare the best-fit JAM model and the mock $V_{\rm rms}$ on the same velocity~scale. 

\begin{table}[H]
    
    \scriptsize
    \caption{Best-fit JAM parameters and their uncertainties for three mock-\Mbh-IFS simulations and five inner power-law slopes of NGC~300.}

    \label{jam_ngc300}

\begin{adjustwidth}{-\extralength}{0cm}

\setlength{\cellWidtha}{\fulllength/11-2\tabcolsep}
\setlength{\cellWidthb}{\fulllength/11-2\tabcolsep}
\setlength{\cellWidthc}{\fulllength/11-2\tabcolsep}
\setlength{\cellWidthd}{\fulllength/11-2\tabcolsep}
\setlength{\cellWidthe}{\fulllength/11-2\tabcolsep}
\setlength{\cellWidthf}{\fulllength/11-2\tabcolsep}
\setlength{\cellWidthg}{\fulllength/11-2\tabcolsep}
\setlength{\cellWidthh}{\fulllength/11-2\tabcolsep}
\setlength{\cellWidthi}{\fulllength/11-2\tabcolsep}
\setlength{\cellWidthj}{\fulllength/11-2\tabcolsep}
\setlength{\cellWidthk}{\fulllength/11-2\tabcolsep}

\begin{tabularx}{\fulllength}{
  >{\centering\arraybackslash}m{\cellWidtha}
  >{\centering\arraybackslash}m{\cellWidthb}
  >{\centering\arraybackslash}m{\cellWidthc}
  >{\centering\arraybackslash}m{\cellWidthd}
  >{\centering\arraybackslash}m{\cellWidthe}
  >{\centering\arraybackslash}m{\cellWidthf}
  >{\centering\arraybackslash}m{\cellWidthg}
  >{\centering\arraybackslash}m{\cellWidthh}
  >{\centering\arraybackslash}m{\cellWidthi}
  >{\centering\arraybackslash}m{\cellWidthj}
  >{\centering\arraybackslash}m{\cellWidthk}
}
\toprule
\textbf{Slope} & \textbf{Parameters} & \multicolumn{3}{c}{\boldmath\textbf{Input $M_{\rm BH} = 0$~M$_\odot$}} & \multicolumn{3}{c}{\boldmath\textbf{Input $\lg(M_{\rm BH}/$M$_\odot) = 3.7$}} & \multicolumn{3}{c}{\boldmath\textbf{Input $\lg(M_{\rm  BH}/$M$_\odot) = 4$}} \\ 
\midrule
\textbf{({Inner} 
 Power Law)} & \textbf{(JAM)} & \textbf{Best-Fit Model} & \boldmath\textbf{1$\sigma$ Error   (16--84\%)}  & \boldmath\textbf{3$\sigma$ Error (0.14--99.86\%)} & \textbf{Best-Fit Model} & \boldmath\textbf{1$\sigma$ Error (16--84\%)} & \textbf{3$\sigma$ Error (0.14--99.86\%) }& \textbf{Best-Fit Model} & \boldmath\textbf{1$\sigma$ Error (16--84\%)} &\boldmath \textbf{3$\sigma$ Error (0.14--99.86\%)} \\ 

\textbf{(1)} & \textbf{(2) }& \textbf{(3)} & \textbf{(4)} & \textbf{(5)} & \textbf{(6)} & \textbf{(7)} & \textbf{(8)} & \textbf{(9)} & \textbf{(10)} & \textbf{(11)} \\ 
\midrule
{$\gamma=0$} 
  & $\lg$(\Mbh/\Msun)         & 0.007  & $\pm0.23$ & $\pm0.5$   & 0.056 & $\pm0.35$ & $\pm0.7$   & 3.916 & $\pm0.007$ & $\pm0.018$ \\ 
            & \ml$_{I}$ (\Msun/\Lsun)   & 0.667  & $\pm0.001$& $\pm0.002$ & 0.682 & $\pm0.002$& $\pm0.003$ & 0.681 & $\pm0.001$ & $\pm0.003$ \\ 
            & $i (^{\circ})$            & 85.35  & $\pm20$   & $\pm32$    & 71.89 & $\pm17$   & $\pm28$    & 80.0  & $\pm8.2$   & $\pm17.2$ \\ 
            & $\beta_z$                 & $-$0.039&$\pm0.008$&$\pm0.024$ & $-$0.020&$\pm0.008$&$\pm0.027$ & $-$0.010&$\pm0.008$&$\pm0.019$ \\ 
\midrule
$\gamma=0.1$& $\lg$(\Mbh/\Msun)         & 0.01   & $\pm0.26$ & $\pm0.58$  & 3.639 & $\pm0.1$  & $\pm0.37$  & 3.978 & $\pm0.007$ & $\pm0.022$ \\ 
            & \ml$_{I}$ (\Msun/\Lsun)   & 0.654  & $\pm0.001$& $\pm0.005$ & 0.639 & $\pm0.001$& $\pm0.002$ & 0.681 & $\pm0.002$ & $\pm0.005$ \\ 
            & $i (^{\circ})$            & 79.34  & $\pm13$   & $\pm23$    & 84.87 & $\pm21$   & $\pm33$    & 74.61 & $\pm20$    & $\pm30$ \\ 
            & $\beta_z$                 & $-$0.026&$\pm0.005$&$\pm0.015$ & $-$0.028&$\pm0.008$&$\pm0.025$ & $-$0.011&$\pm0.008$&$\pm0.018$ \\ 
\midrule
$\gamma=0.2$& $\lg$(\Mbh/\Msun)         & 0.009  & $\pm0.39$ & $\pm0.7$   & 3.816 & $\pm0.014$& $\pm0.037$ & 4.052 & $\pm0.004$ & $\pm0.011$ \\ 
            & \ml$_{I}$ (\Msun/\Lsun)   & 0.640  & $\pm0.002$& $\pm0.005$ & 0.617 & $\pm0.001$& $\pm0.002$ & 0.647 & $\pm0.002$ & $\pm0.003$ \\ 
            & $i (^{\circ})$            & 78.36  & $\pm13$   & $\pm24$    & 84.5  & $\pm16$   & $\pm29$    & 88.73 & $\pm19$    & $\pm30$ \\ 
            & $\beta_z$                 & $-$0.016&$\pm0.003$&$\pm0.011$ & $-$0.013&$\pm0.005$&$\pm0.02$  & $-$0.011&$\pm0.004$&$\pm0.017$ \\ 
\midrule
$\gamma=0.5$& $\lg$(\Mbh/\Msun)         & 0.017  & $\pm0.36$ & $\pm0.8$   & 3.717 & $\pm0.003$& $\pm0.01$  & 3.892 & $\pm0.002$ & $\pm0.008$ \\ 
            & \ml$_{I}$ (\Msun/\Lsun)   & 0.618  & $\pm0.001$& $\pm0.003$ & 0.560 & $\pm0.002$& $\pm0.005$ & 0.636 & $\pm0.001$ & $\pm0.007$ \\ 
            & $i (^{\circ})$            & 80.43  & $\pm11$   & $\pm21$    & 79.98 & $\pm3.8$  & $\pm12$    & 80.1  & $\pm11$    & $\pm22$ \\ 
            & $\beta_z$                 & $-$0.024&$\pm0.003$&$\pm0.011$ & $-$0.03 &$\pm0.001$&$\pm0.004$ & $-$0.017&$\pm0.002$&$\pm0.009$ \\ 
\midrule 
$\gamma=0.7$& $\lg$(\Mbh/\Msun)         & 0.001  & $\pm0.033$& $\pm0.1$   & 0.001 & $\pm0.012$& $\pm0.047$ & 0.005 & $\pm0.059$ & $\pm0.17$ \\ 
            & \ml$_{I}$ (\Msun/\Lsun)   & 0.427  & $\pm0.001$& $\pm0.002$ & 0.444 & $\pm0.001$& $\pm0.003$ & 0.465 & $\pm0.002$ & $\pm0.003$ \\ 
            & $i (^{\circ})$            & 77.28  & $\pm2.2$  & $\pm7.1$   & 77.78 & $\pm1.4$  & $\pm3.8$   & 88.04 & $\pm4.4$  & $\pm9.9$ \\ 
            & $\beta_z$                 & $-$0.090&$\pm0.002$&$\pm0.007$ & $-$0.076&$\pm0.001$&$\pm0.003$ & $-$0.058&$\pm0.002$&$\pm0.005$ \\ 
\bottomrule 
\end{tabularx}

\end{adjustwidth}

\noindent{\footnotesize{\textit{Notes:} Column 1: The inner power-law slopes of the core-S\'ersic function of NSC. Column 2: The JAM’s parameters. Columns 3--5: The best-fit parameters associated with the case of no BH ($M_{\rm BH} = 0$~M$_\odot$), 1$\sigma$ (or 16--84\%), and~3$\sigma$ (or 0.14--99.86\%) uncertainties provided by the {\tt adamet} MCMC with JAM, respectively. Columns 6--8: Similarities of Columns 3--5 but for the case of input $M_{\rm BH}=5\times10^3$ M$_\odot$. Columns 9--11: Similarities of Columns 3--5 but for the case of input $M_{\rm BH}=10^4$ M$_\odot$. Modeling summary in this Table is associated with Figures~\ref{new_ngc300_gamma_0.0}--\ref{new_ngc300_gamma_0.7}.}}

\end{table}

One notable feature of the stellar velocity dispersion produced by the best-fit JAM is clearly illustrated in Figures~\ref{new_ngc300_gamma_0.0}--\ref{new_ngc300_gamma_0.7} for NGC~300, in~the case with no BH. However, this pattern is less evident in the corresponding models for NGC~3115~dw01, as~shown in Figures~\ref{new3_ngc3115dw01_gamma_0.0}--\ref{new3_ngc3115dw01_gamma_0.7}. As~the $\gamma$ value increases, the~central velocity dispersion drop becomes more pronounced, with~a larger extent and greater depth, requiring a larger $M_{\rm BH}$ for JAM to reproduce the mock $V_{\rm rms}$. An~increase in $\gamma$ also raises the central stellar mass density, and~JAM utilizes the \ml\ parameter to account for the degeneracy between stellar mass density and \Mbh. We explore these complexities to explain our results in detail~below.

\begin{table}[H]

    \scriptsize
    \caption{Best-fit JAM parameters and their uncertainties for three mock-\Mbh-IFS simulations and five inner power-law slopes of NGC~3115~dw01.}

    \label{jam_ngc3115dw01}

\begin{adjustwidth}{-\extralength}{0cm}

\setlength{\cellWidtha}{\fulllength/11-2\tabcolsep}
\setlength{\cellWidthb}{\fulllength/11-2\tabcolsep}
\setlength{\cellWidthc}{\fulllength/11-2\tabcolsep}
\setlength{\cellWidthd}{\fulllength/11-2\tabcolsep}
\setlength{\cellWidthe}{\dimexpr\fulllength/11 - 2\tabcolsep\relax}
\setlength{\cellWidthf}{\fulllength/11-2\tabcolsep}
\setlength{\cellWidthg}{\fulllength/11-2\tabcolsep}
\setlength{\cellWidthh}{\fulllength/11-2\tabcolsep}
\setlength{\cellWidthi}{\fulllength/11-2\tabcolsep}
\setlength{\cellWidthj}{\fulllength/11-2\tabcolsep}
\setlength{\cellWidthk}{\fulllength/11-2\tabcolsep}

\begin{tabularx}{\fulllength}{
  >{\centering\arraybackslash}m{\cellWidtha}
  >{\centering\arraybackslash}m{\cellWidthb}
  >{\centering\arraybackslash}m{\cellWidthc}
  >{\centering\arraybackslash}m{\cellWidthd}
  >{\centering\arraybackslash}m{\cellWidthe}
  >{\centering\arraybackslash}m{\cellWidthf}
  >{\centering\arraybackslash}m{\cellWidthg}
  >{\centering\arraybackslash}m{\cellWidthh}
  >{\centering\arraybackslash}m{\cellWidthi}
  >{\centering\arraybackslash}m{\cellWidthj}
  >{\centering\arraybackslash}m{\cellWidthk}
}
\toprule
\textbf{Slope} & \textbf{Parameters} & \multicolumn{3}{c}{\boldmath\textbf{Input $M_{\rm BH} = 0$~M$_\odot$}} & \multicolumn{3}{c}{\boldmath\textbf{Input $\lg(M_{\rm BH}/$M$_\odot) = 4.544$}} & \multicolumn{3}{c}{\boldmath\textbf{Input $\lg(M_{\rm  BH}/$M$_\odot) = 4.845$}} \\ 
\midrule
\textbf{(Inner Power Law) }& \textbf{(JAM)} & \textbf{Best-Fit Model }& \boldmath\textbf{1$\sigma$ Error (16--84\%)} & \boldmath\textbf{3$\sigma$ Error (0.14--99.86\%) }& \textbf{Best-Fit Model} &\boldmath \textbf{1$\sigma$ Error (16--84\%)} & \boldmath\textbf{3$\sigma$ Error (0.14--99.86\%)} & \textbf{Best-Fit Model} &\boldmath\textbf{ 1$\sigma$ Error (16--84\%)} & \boldmath\textbf{3$\sigma$ Error (0.14--99.86\%)} \\ 
\textbf{(1)} & \textbf{(2)} & \textbf{(3)} & \textbf{(4)} & \textbf{(5)} & \textbf{(6)} & \textbf{(7)} & \textbf{(8)} & \textbf{(9)} & \textbf{(10)} & \textbf{(11)} \\ 
\midrule
$\gamma=0$   & $\lg$(\Mbh/\Msun)       & 0.181 & $\pm0.75$ & $\pm1.4$  & 4.14  & $\pm0.01$ & $\pm0.031$ & 4.321 & $\pm0.005$ & $\pm0.016$ \\ 
             & \ml$_{I}$ (\Msun/\Lsun) & 1.569 & $\pm0.001$& $\pm0.002$& 1.56  & $\pm0.001$& $\pm0.002$ & 1.556 & $\pm0.001$ & $\pm0.002$ \\ 
             & $i (^{\circ})$          & 30    & $\pm0.019$& $\pm0.069$& 39.25 & $\pm0.27$ & $\pm0.75$  & 76.05 & $\pm5.9$   & $\pm10$ \\ 
             & $\beta_z$               & 0.1& $\pm0.002$&$\pm0.007$& 0.068&$\pm0.001$&$\pm0.003$ & 0.026&$\pm0.001$&$\pm0.002$ \\ 
\midrule
$\gamma=0.1$ & $\lg$(\Mbh/\Msun)       & 0.393 & $\pm0.93$ & $\pm1.6$  & 4.331 & $\pm0.021$& $\pm0.058$ & 4.875 & $\pm0.006$ & $\pm0.01$ \\ 
             & \ml$_{I}$ (\Msun/\Lsun) & 1.472 & $\pm0.001$& $\pm0.003$& 1.475 & $\pm0.001$& $\pm0.002$ & 1.481 & $\pm0.001$ & $\pm0.002$\\ 
             & $i (^{\circ})$          & 87.35 & $\pm20$   & $\pm30$   & 81.01 & $\pm9.4$  & $\pm20$    & 79.99 & $\pm12$    & $\pm21$\\ 
             & $\beta_z$               & 0.001&$\pm0.001$&$\pm0.006$& 0.004&$\pm0.001$&$\pm0.002$& 0.004&$\pm0.001$&$\pm0.002$  \\ 
\midrule
$\gamma=0.2$ & $\lg$(\Mbh/\Msun)       & 0.374 & $\pm0.78$ & $\pm1.3$  & 4.506 & $\pm0.013$& $\pm0.043$ & 4.81  & $\pm0.007$ & $\pm0.02$ \\ 
             & \ml$_{I}$ (\Msun/\Lsun) & 1.454 & $\pm0.001$& $\pm0.002$& 1.455 & $\pm0.001$& $\pm0.002$ & 1.503 & $\pm0.001$ & $\pm0.003$ \\ 
             & $i (^{\circ})$          & 74.87 & $\pm13$   & $\pm24$   & 77.68 & $\pm9.9$  & $\pm20$    & 87.94 & $\pm11$    & $\pm23$ \\ 
             & $\beta_z$               & 0.005&$\pm0.001$&$\pm0.004$& 0.007&$\pm0.001$&$\pm0.003$& 0.0006&$\pm0.0004$&$\pm0.0001$ \\ 
\midrule
$\gamma=0.4$ & $\lg$(\Mbh/\Msun)       & 0.08  & $\pm0.45$ & $\pm0.92$ & 0.127 & $\pm0.55$ & $\pm1.1$   & 4.208 & $\pm0.069$ & $\pm0.21$ \\ 
             & \ml$_{I}$ (\Msun/\Lsun) & 1.121 & $\pm0.001$& $\pm0.002$& 1.133 & $\pm0.001$& $\pm0.003$ & 1.16  & $\pm0.001$ & $\pm0.003$ \\ 
             & $i (^{\circ})$          & 82.77 & $\pm7.9$  & $\pm17$   & 88.81 & $\pm6.6$  & $\pm15$    & 86.43 & $\pm20$    & $\pm30$ \\ 
             & $\beta_z$               & 0.002&$\pm0.001$&$\pm0.004$& 0.001&$\pm0.001$&$\pm0.002$& 0.001&$\pm0.007$&$\pm0.02$ \\ 
\midrule 
$\gamma=0.7$ & $\lg$(\Mbh/\Msun)       & 0.042 & $\pm0.29$& $\pm0.61$ & 0.026 & $\pm0.17$ & $\pm0.43$  & 0.135 & $\pm0.44$  & $\pm0.83$ \\ 
             & \ml$_{I}$ (\Msun/\Lsun) & 0.830 & $\pm0.001$& $\pm0.002$& 0.841 & $\pm0.001$& $\pm0.002$ & 0.868 & $\pm0.001$ & $\pm0.002$ \\ 
             & $i (^{\circ})$          & 79.98 & $\pm5.6$ & $\pm16$   & 88.33 & $\pm3.1$  & $\pm6.7$   & 80.81 & $\pm9.3$   & $\pm19$ \\ 
             & $\beta_z$               & 0.004&$\pm0.003$&$\pm0.012$& 0.007&$\pm0.001$&$\pm0.002$& 0.006&$\pm0.005$&$\pm0.016$ \\ 
\bottomrule 
\end{tabularx}
\end{adjustwidth}

\noindent{\footnotesize{\textit{Notes:} Columns 1--5: Same as Table~\ref{jam_ngc300}. Columns 6--8: Similarities of Columns 3--5 but for the case of input $M_{\rm BH}=3.5\times10^4$ M$_\odot$. Columns 9--11: Similarities of Columns 3--5 but for the case of input $M_{\rm BH}=7\times10^4$ M$_\odot$. Modeling summary in this Table is associated with Figures~\ref{new3_ngc3115dw01_gamma_0.0}--\ref{new3_ngc3115dw01_gamma_0.7}.} }

\end{table}

For the tests with $\gamma=0$ ({\it flat core}, see the top-row panels of Figures~\ref{new_ngc300_gamma_0.0}--\ref{new_ngc300_gamma_0.7} for NGC~300 and Figures~\ref{new3_ngc3115dw01_gamma_0.0}--\ref{new3_ngc3115dw01_gamma_0.7} for NGC~3115~dw01), JAM cannot replicate the central drop in velocity dispersion (and subsequently in the $V_{\rm rms}$ maps) for cases with input $M_{\rm BH}=0$ \Msun. This is because the projected dispersion predicted by JAM remains flat towards the centers when $\gamma=0$, e.g.,~see {Figure~4}  
of {N25} and \citet{Tremaine94}, leading to biased lower $M_{\rm BH}$ estimates and resulting in an upper limit (NGC~300) or lower $M_{\rm BH}$ estimates (NGC~3115~dw01) for the mock kinematic cases with input $M_{\rm BH}=0.5\%$ or $1\%$ \Mnsc. In~these scenarios, there is no cusp in the SB, which is dominated by stars with radii around the break radius, and~the density follows $\rho \propto r^{-1}$. Consequently, when there is no BH, the~depth of the central potential well is finite, and~the projected dispersion remains asymptotically constant as $r \rightarrow 0$, dominated by stars inside the break radius. If~a BH is present, the~projected dispersion increases at a rate between $\sigma_\star \propto r^{-1}$ and $r^{-0.5}$.

In all tests with the {\it weak cusp} models ($0 < \gamma \leq 0.2$), we successfully reproduced the results from {N25} with $\gamma = 0.1$, achieving uncertainties within 30\% for \ml$_{I}$ and 10\% for \Mbh. This outcome was expected, as~the MICADO images were modeled using the HST SB constrained in {N25}. Additionally, we validated the numerical prediction for $\gamma = 0.2$ in NGC~300, as~reported by {N25}, by~replicating it with our mock MICADO simulations. This further confirms the accuracy of the same $\gamma$ simulations for NGC~3115~dw01 in this~work.

However, in~the case of NGC~300, there is an opposite trend in \ml$_{I}$ and \Mbh\ compared to their trends in {N25}. Specifically, our JAM with the MICADO mass models provided lower \Mbh\ but higher \ml$_{I}$ than those in {N25} for the same mock kinematics. This discrepancy is due to the MICADO images providing more light (and thus stellar mass) in the radius range of 0$\farcs$2--3$\farcs$0 than the best-fit surface-brightness profiles constrained from HST images, as~seen in Figure~\ref{mass-model_NGC300}. In~the case of NGC~3115~dw01, the~trend of \ml$_{I}$ and \Mbh\ is entirely consistent with the findings from {N25}, due to the consistency between its MICADO and HST images (see Figure~\ref{mass-model_NGC3115dw01}). Here, the~dynamical model compensates for the central velocity dispersion drop primarily through the negative covariance between \Mbh\ and \ml$_{I}$. The~increase in central stellar mass density remains either lower than or comparable to the \Mbh, thus having a negligible impact on \Mbh\ determinations. Specifically, for~NGC~300, the~difference in central stellar mass is $M_{\star,\gamma=0.2} - M_{\star,\gamma=0} \approx 1.4 \times 10^4$ \Msun, and~for NGC~3115~dw01, it is $M_{\star,\gamma=0.2} - M_{\star,\gamma=0} \approx 7.4 \times 10^4$ \Msun.

The 2D PDFs show the 3$\sigma$ confidence level “banana shape” of the negative covariance between $M_{\rm BH}$ and \ml$_I$. This covariance occurs due to the interplay between the gravitational potentials of central BHs and their host galaxies probed by both MICADO and HARMONI, where larger \Mbh\ correspond to smaller \ml$_I$, and~vice~versa.

For the {\it presumably intermediate cusp} models ($0.2 < \gamma \leq 0.5$), we considered $\gamma=0.5$ for NGC~300 and $\gamma=0.4$ for NGC~3115~dw01. In~these cases, the~stellar mass density difference between these $\gamma$ profiles and the profile with $\gamma=0$ becomes comparable to, or~even exceeds, the~BH mass  within $0\farcs1$ of the core-Sérsic profile of the NSCs. Specifically, $M_{\star,\gamma=0.5} - M_{\star,\gamma=0} \approx 4.25 \times 10^4$ \Msun\ for NGC~300 (Figure~\ref{mass-model_NGC300}) and $M_{\star,\gamma=0.4} - M_{\star,\gamma=0} \approx 7.73 \times 10^5$ \Msun\ for NGC~3115~dw01 (Figure~\ref{mass-model_NGC3115dw01}), leading JAM  to predict alower \Mbh\ or fail to detect it (see Figures~\ref{new_ngc300_gamma_0.0}--\ref{new_ngc300_gamma_0.7} {and} 
 Figures~\ref{new3_ngc3115dw01_gamma_0.0}--\ref{new3_ngc3115dw01_gamma_0.7}).

For the {\it strong cusp} models ($\gamma > 0.5$), we considered $\gamma=0.7$ for both NGC~300 and  NGC~3115~dw01. In~these models, the~central stellar mass density of the NSC significantly surpasses the \Mbh. For~NGC~300, the~increase in stellar mass is $M_{\star,\gamma=0.7} - M_{\star,\gamma=0} \approx 2.0 \times 10^6$ \Msun\ (Figure~\ref{mass-model_NGC300}), and~for NGC~3115~dw01, it is $M_{\star,\gamma=0.7} - M_{\star,\gamma=0} \approx 1.55 \times 10^6$ \Msun\ (Figure~\ref{mass-model_NGC3115dw01}). As~a result, the~\Mbh\ is completely~diminished.

In addition, it should also be noted that the variation of $\gamma$ does not impact the constraints on the inclination ($i$), which remains weakly constrained in our dynamical models due to the round shape of NSCs. Similarly, the~orbital anisotropy ($\beta_z$) is only slightly negative for NGC~300 and positive for NGC~3115~dw01 but close to zero, the~assumed value in the {N25} HSIM simulations, indicating minimal~effect. 

Another note is that a central cusp in an NSC without a black hole can arise from collisional stellar dynamics, in~which two-body relaxation drives mass segregation and core contraction \citep{Bahcall77, Valluri05, Nguyen2025b}, leading to a steepened central density profile. Additional contributors include dissipative gas inflows followed by centrally concentrated star formation and~the inspiral and merging of dense star clusters via dynamical friction, both of which can build a cuspy stellar distribution independent of a central black~hole.

\section{Conclusions}\label{Conclusion}

We have presented novel ELT/MICADO simulations of galaxy + NSC stellar kinematics seeking to constrain IMBHs using {N25}'s mock HARMONI kinematics. We summarize our conclusions as {follows:} 

\begin{enumerate}
\item[(i)] We enhanced the SimCADO software, used for simulating MICADO images on the ELT, by~incorporating a core-Sérsic function to model the SB of~NSCs. 

\item[(ii) ] Our mock MICADO $I$-band images for NGC~300 and NGC~3115~dw01, assuming an inner power-law slope of $\gamma = 0.1$ for the core-Sérsic profile, are fully consistent with the constraints from {N25}. This holds true at both the larger scales observed with HST and the interpolated 4 mas scale. Given this consistency for $\gamma = 0.1$, we used it as a reference for further simulations with slopes of $\gamma = 0,\, 0.2,\, 0.5,\, 0.7$ for NGC~300, and~$\gamma = 0,\, 0.2,\, 0.4,\, 0.7$ for NGC~3115~dw01. 

\item[(iii)]  These mock MICADO images (4-mas pixel size) for both galaxies can be achieved with a one-hour on-source exposure time. Since their SBs remain consistent with the existing HST observational scale at large radii, increasing the exposure time to four hours did not result in significant image quality improvement. Therefore, we set one hour as the maximum on-source exposure time for MICADO observations in the HARMONI IMBH~sample. 

\item[(iv)]  We used mock MICADO images to reconstruct the stellar mass models for both galaxies. These models, combined with mock HARMONI kinematics from {N25}, served as inputs for JAM to re-estimate the masses of their central IMBHs and measurement uncertainties. For~models assuming $\gamma = 0$ mass profiles (i.e., flat-core NSCs), the~best-fitting JAM solutions—independent of the Sérsic index—recover a nearly constant or rising central velocity dispersion, rather than the central dispersion drop expected for systems hosting zero or very low-mass IMBHs [{N25}]~\citep{Tremaine94}. This behavior leads to a systematic underestimation of the inferred IMBH~mass.

\item[(v)]  Our same tests with assumed $\gamma > 0$ mass profiles, and~increasing $\gamma$ for the NSCs, exaggerate the central drop in both the size and depth of the velocity dispersion at a few central spaxels, where the stellar kinematics are dominated by the gravitational potential of the IMBH. This feature was also shown in {Figure~4} of {N25} 
 and discussed in {Section~4.2} of {N25}, though~based on numerical predictions. Here, we confirmed the $\gamma$-dependence of stellar kinematics by simulating realistic photometric observations using MICADO, in~combination with mock HSIM IFS HARMONI kinematics and JAM. While increasing $\gamma$ causes significant drops in central velocity dispersion, {\it it also raises the stellar mass density in that region (within $0\farcs1$) compared to that of the {\it flat core} models}, which is accounted for and balanced in the dynamical~models.
\begin{itemize}
\item {\it Weak cusp} ($0 < \gamma \leq 0.2$): While the central stellar mass density increases, it remains lower than or comparable to the \Mbh. Thus, JAM compensates for the drop in central velocity dispersion primarily through the negative covariance between \Mbh\ and \ml.     

\item {\it Presumably intermediate cusp} ($0.2 < \gamma \leq 0.5$): The central stellar mass density increases significantly and even surpasses the \Mbh. JAM thus tends to underestimate the IMBH mass or yields no~detection.

\item {\it Strong cusp} ($\gamma > 0.5$): The central stellar mass density becomes dominant over the \Mbh.  As~a result, the~BH \Mbh\ is entirely suppressed in~JAM.

\end{itemize}
\end{enumerate}

{{Software:} 
}{ {\tt Python~3.12:} \citep{VanRossum2009}; 
{\tt Matplotlib~3.6:} \citep{Hunter2007};
{\tt NumPy~1.22:} \citep{Harris2020}; 
{\tt SciPy~1.3:} \citep{Virtanen2020};  
{\tt photutils~0.7:} \citep{bradley2024}; 
{\tt AstroPy~5.1} \citep{AstropyCollaboration22}; 
{\tt AdaMet 2.0} \citep{Cappellari13a}; 
{\tt JamPy~7.2} \citep{Cappellari20}; 
{\tt MgeFit~5.0} \citep{Cappellari02}; and~SimCADO \citep{Leschinski16}.
}

\vspace{6pt}
\authorcontributions{
Conceptualization, Tinh Q. T. Le, Dieu D. Nguyen and Hai N. Ngo; 
Methodology, Tinh Q. T. Le, Dieu D. Nguyen and Hai N. Ngo; 
Software, Tinh Q. T. Le, Dieu D. Nguyen and Hai N. Ngo; 
Validation, Tinh Q. T. Le, Hai N. Ngo, Tien H. T. Ho, Tuan N. Le 
and Long Q. T. Nguyen; 
Formal analysis, Hai N. Ngo, Tien H. T. Ho, Tuan N. Le 
and Long Q. T. Nguyen; 
Investigation, Tinh Q. T. Le, Dieu D. Nguyen, Hai N. Ngo, 
Tien H. T. Ho, Tuan N. Le and Long Q. T. Nguyen; 
Resources, Tinh Q. T. Le and Dieu D. Nguyen; 
Data curation, Tinh Q. T. Le, Dieu D. Nguyen, Hai N. Ngo, 
Tien H. T. Ho, Tuan N. Le and Long Q. T. Nguyen; 
Writing---original draft preparation, Dieu D. Nguyen; 
Writing---review and editing, Tinh Q. T. Le and Dieu D. Nguyen; 
Visualization, Tinh Q. T. Le; 
Supervision, Dieu D. Nguyen; 
Funding acquisition, Dieu D. Nguyen. 
All authors have read and agreed to the published version of the manuscript.
}

\funding{\ This research received no external funding. The APC was funded by the authors. 
}

\dataavailability{\ Data generated from this work will be made available for further research upon request to the corresponding author. 
} 


\acknowledgments{
\ 
 The authors would like to thank the anonymous referee for their careful reading and useful comments, which helped to improve the paper~greatly. }

\conflictsofinterest{\ The authors declare no conflicts of interest. 
}

\appendixtitles{yes} 
\appendixstart
\appendix
\section[\appendixname~\thesection]{Supplementary~Figures} 
\begin{figure}[H]
    \includegraphics[width=0.45\textwidth]
    {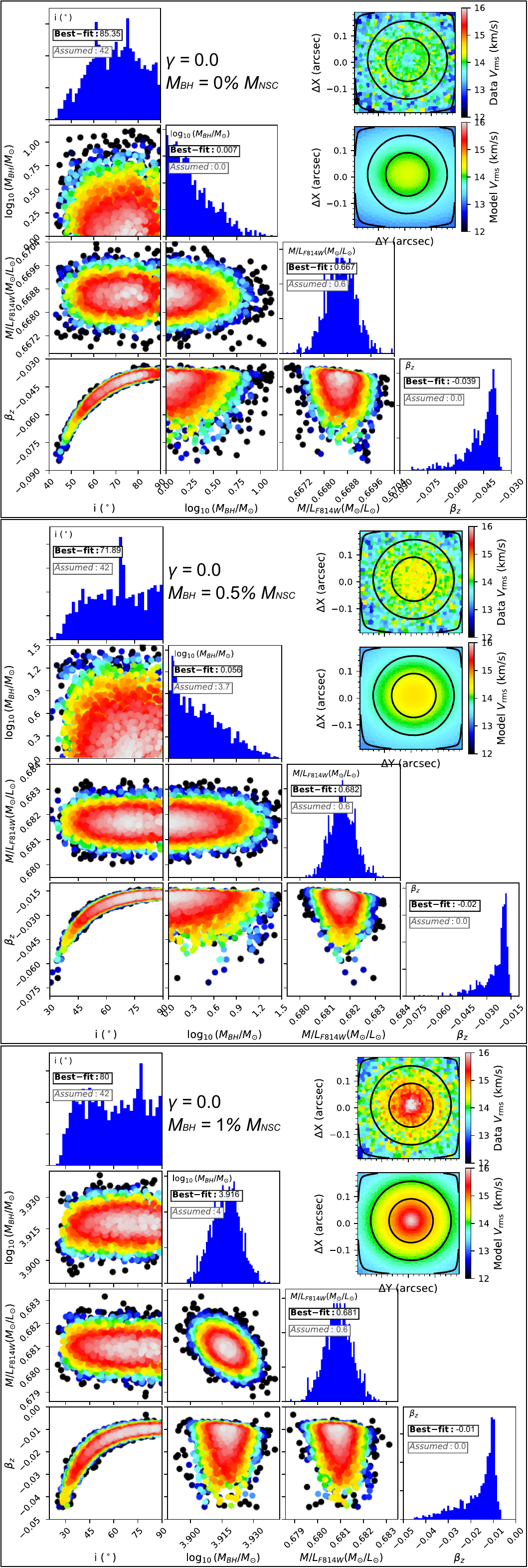}
    \caption{{Posterior} 
 distributions from the {\tt adamet} MCMC fits of JAM models to the mock HSIM \emph{H}-high kinematics and MICADO \emph{I}-band images of NGC~300, for an inner power-law slope $\gamma = 0$. The three rows correspond to three input black hole masses: $M_{\rm BH} = 0~{\rm M}_{\odot}$ {(\textbf{upper}),} 
 $5 \times 10^{3}~{\rm M}_{\odot}$ (\textbf{middle}), and $10^{4}~{\rm M}_{\odot}$ (\textbf{lower}). In each panel, the four free parameters are shown as 2D scatter plots of their projected posteriors and as 1D histograms of their marginalized distributions; white indicates the maximum likelihood and black the $3\sigma$ contour. Inset maps at the top-right of each panel show the mock $V_{\rm rms}$ from N25; the bottom maps show the corresponding $V_{\rm rms}$ recovered by the best-fit JAM model on the same color scale.}
    \label{new_ngc300_gamma_0.0} 
\end{figure}

\begin{figure}[H] 
    \includegraphics[width=0.55\textwidth]
    {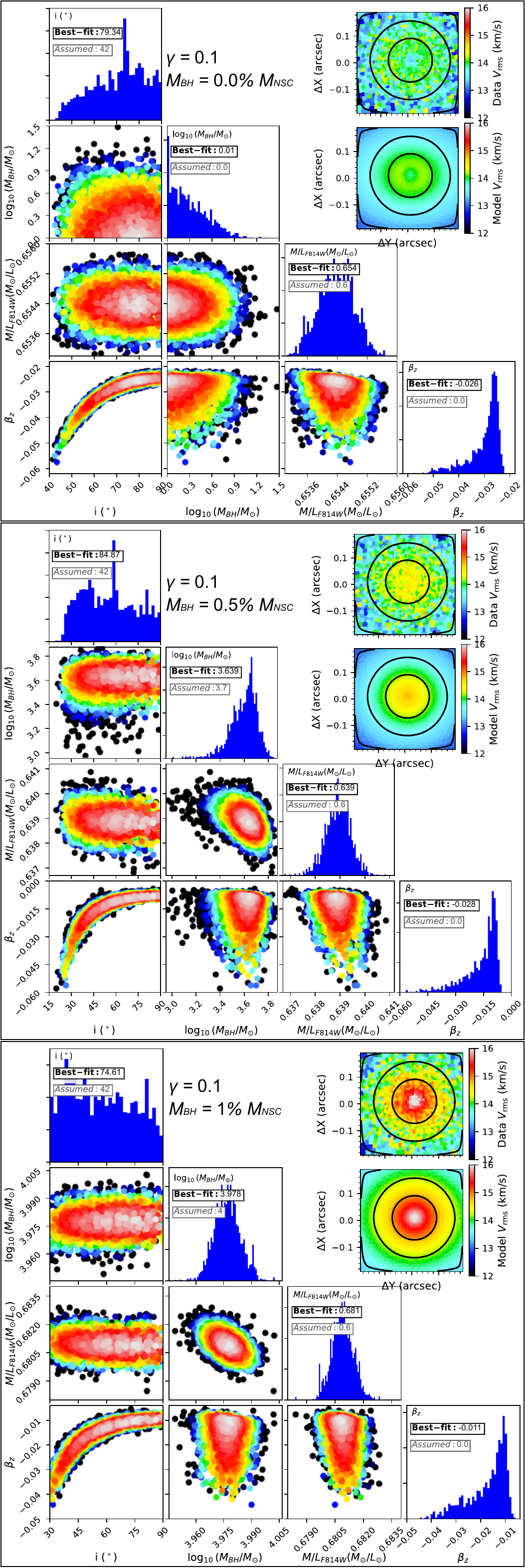}
    \caption{{As} 
 Figure~\ref{new_ngc300_gamma_0.0}, but for $\gamma = 0.1$.}
    \label{new_ngc300_gamma_0.1}
\end{figure}

\begin{figure}[H] 
    \includegraphics[width=0.55\textwidth]
    {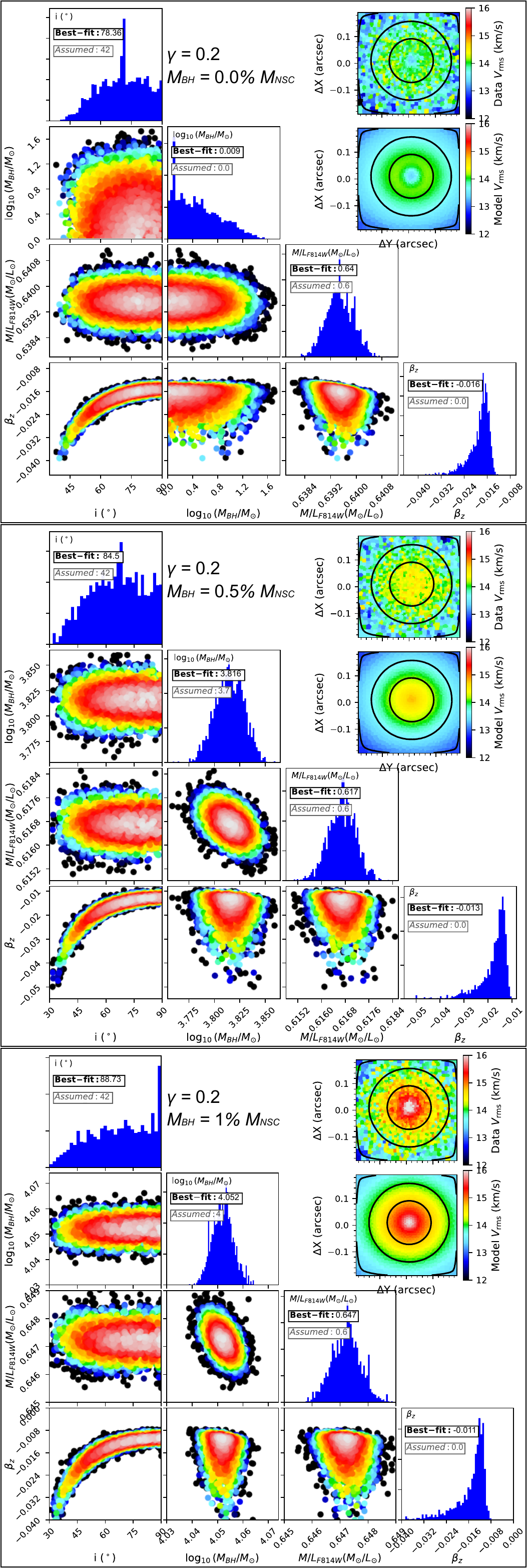}
    \caption{{As} 
 Figure~\ref{new_ngc300_gamma_0.0}, but for $\gamma = 0.2$.}
    \label{new_ngc300_gamma_0.2}
\end{figure}

\begin{figure}[H] 
    \includegraphics[width=0.55\textwidth]
    {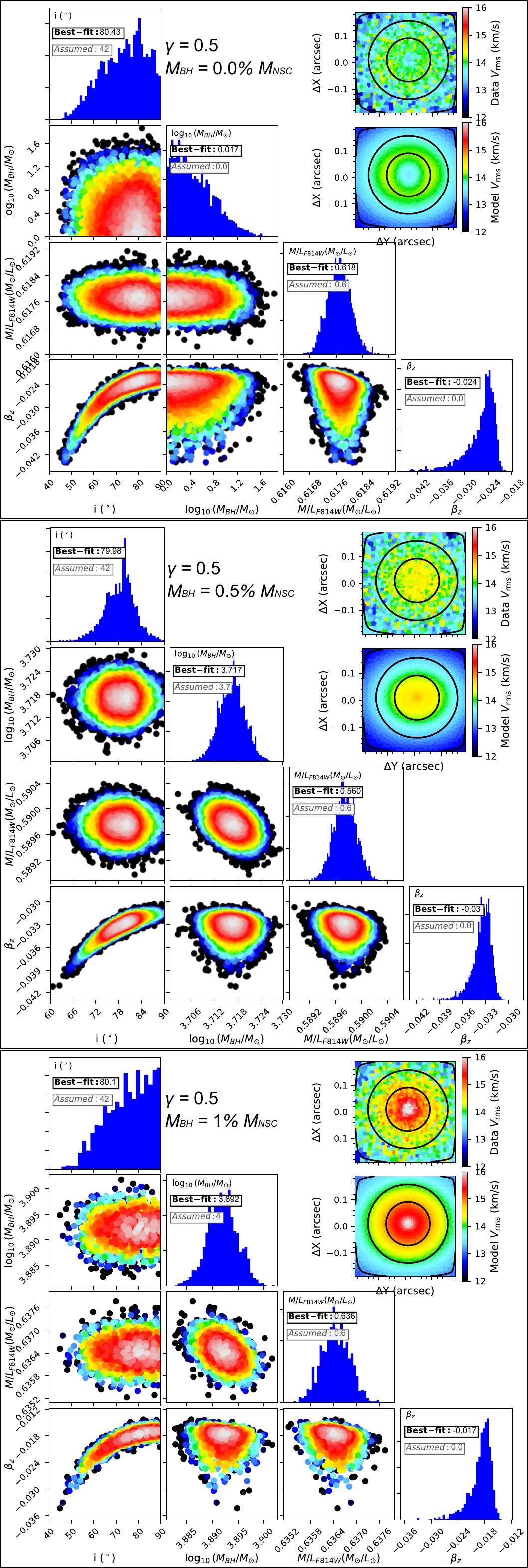}
    \caption{{As} 
  Figure~\ref{new_ngc300_gamma_0.0}, but for $\gamma = 0.5$.}
    \label{new_ngc300_gamma_0.5}
\end{figure}

\begin{figure}[H] 
    
    \includegraphics[width=0.55\textwidth]
    {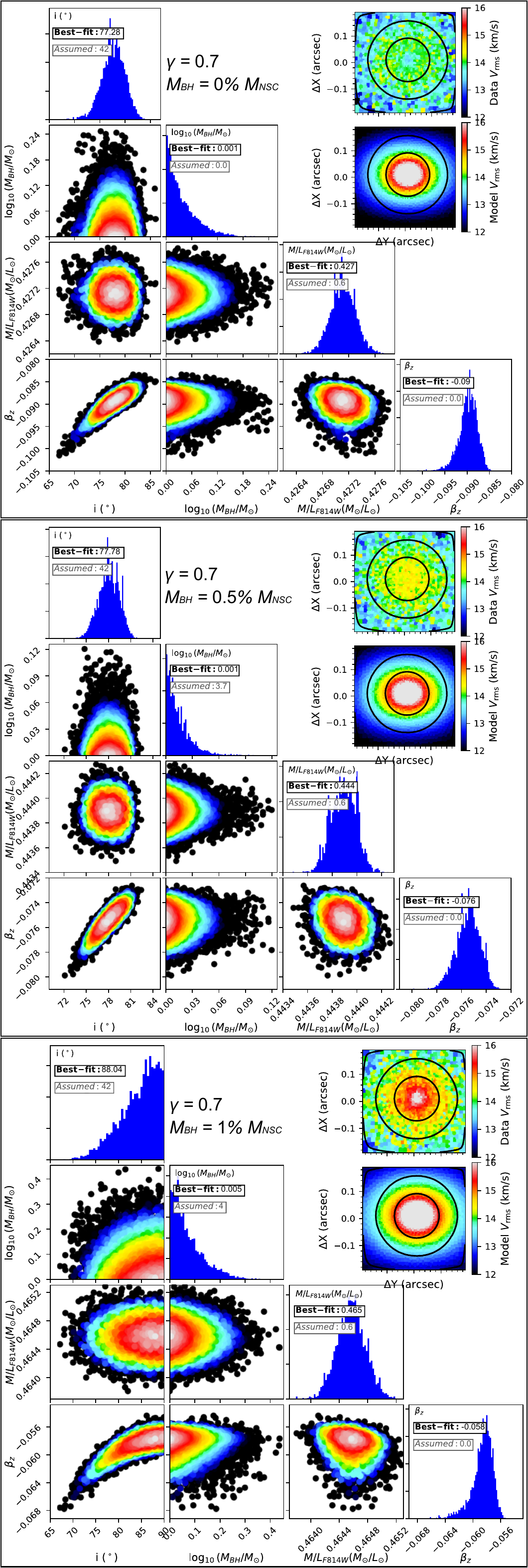}
    \caption{{As} 
  Figure~\ref{new_ngc300_gamma_0.0}, but for $\gamma = 0.7$.}
    \label{new_ngc300_gamma_0.7}
\end{figure}


\begin{figure}[H]
    
    \includegraphics[width=0.52\textwidth]{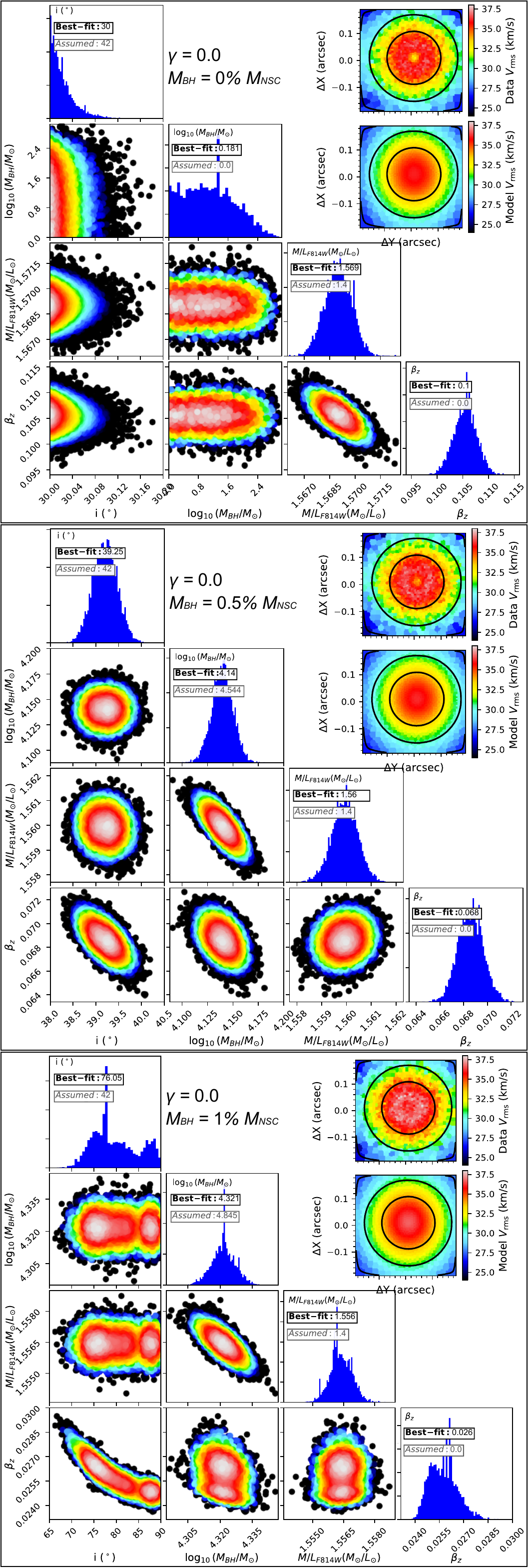}
    \caption{{Posterior} 
 distributions from the {\tt adamet} MCMC fits of JAM models to the mock HSIM \emph{H}-high kinematics and MICADO \emph{I}-band images of NGC~3115~dw01, for an inner power-law slope $\gamma = 0$. The three rows correspond to three input black hole masses: $M_{\rm BH} = 0~{\rm M}_{\odot}$ (\textbf{upper}), $3.5 \times 10^{4}~{\rm M}_{\odot}$ (\textbf{middle}), and $7 \times 10^{4}~{\rm M}_{\odot}$ (\textbf{lower}). The panel layout and color coding are identical to Figure~\ref{new_ngc300_gamma_0.0}.}
    \label{new3_ngc3115dw01_gamma_0.0}
\end{figure}

\begin{figure}[H]
    
    \includegraphics[width=0.55\textwidth]{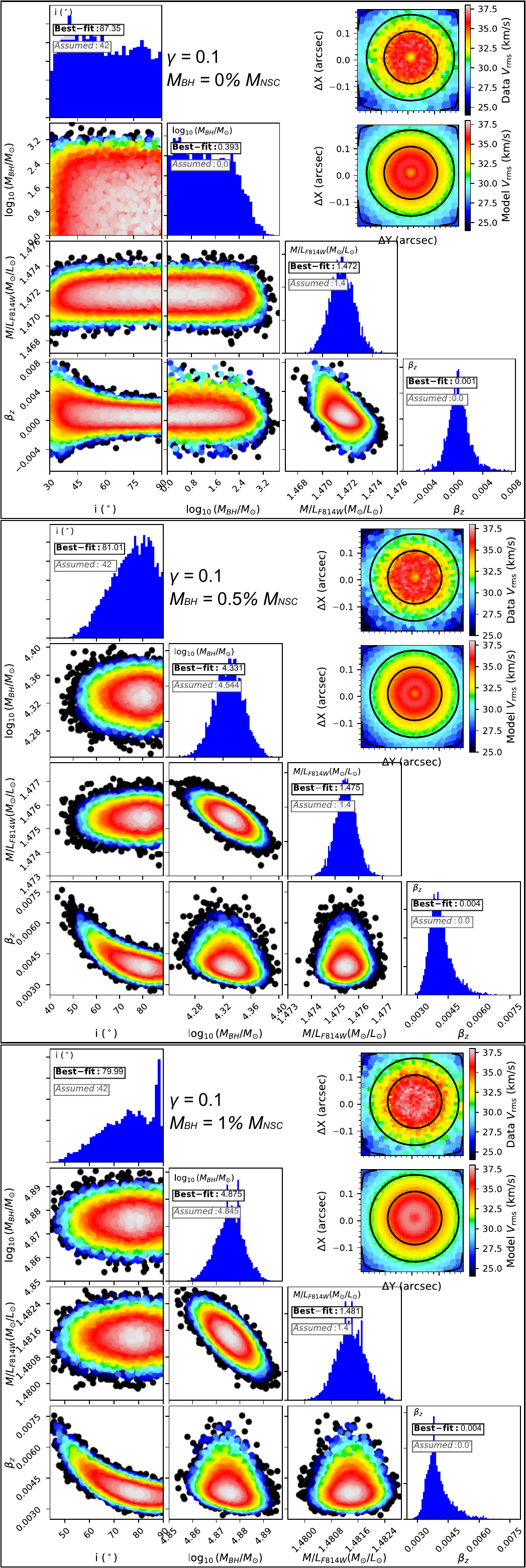}
    \caption{{As} 
 Figure~\ref{new3_ngc3115dw01_gamma_0.0}, but for $\gamma = 0.1$.}
    \label{new3_ngc3115dw01_gamma_0.1}
\end{figure}

\begin{figure}[H]
    
    \includegraphics[width=0.55\textwidth]{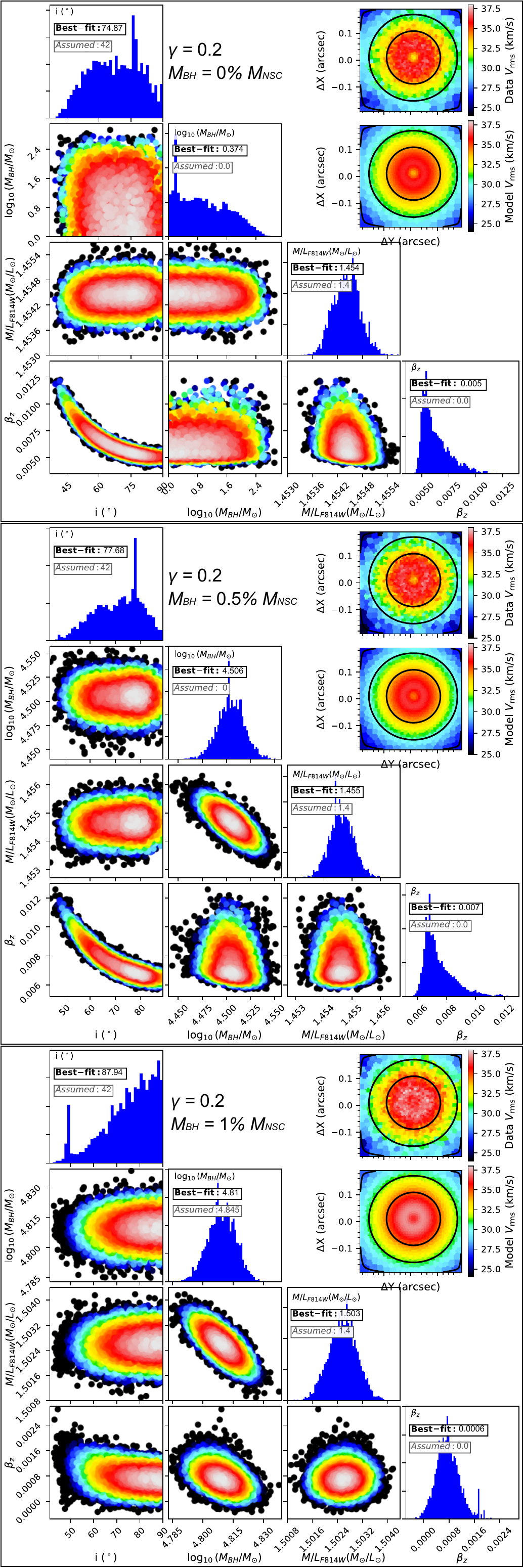}
    \caption{{As} 
 Figure~\ref{new3_ngc3115dw01_gamma_0.0}, but for $\gamma = 0.2$.}
    \label{new3_ngc3115dw01_gamma_0.2}
\end{figure}

\begin{figure}[H]
    
    \includegraphics[width=0.55\textwidth]{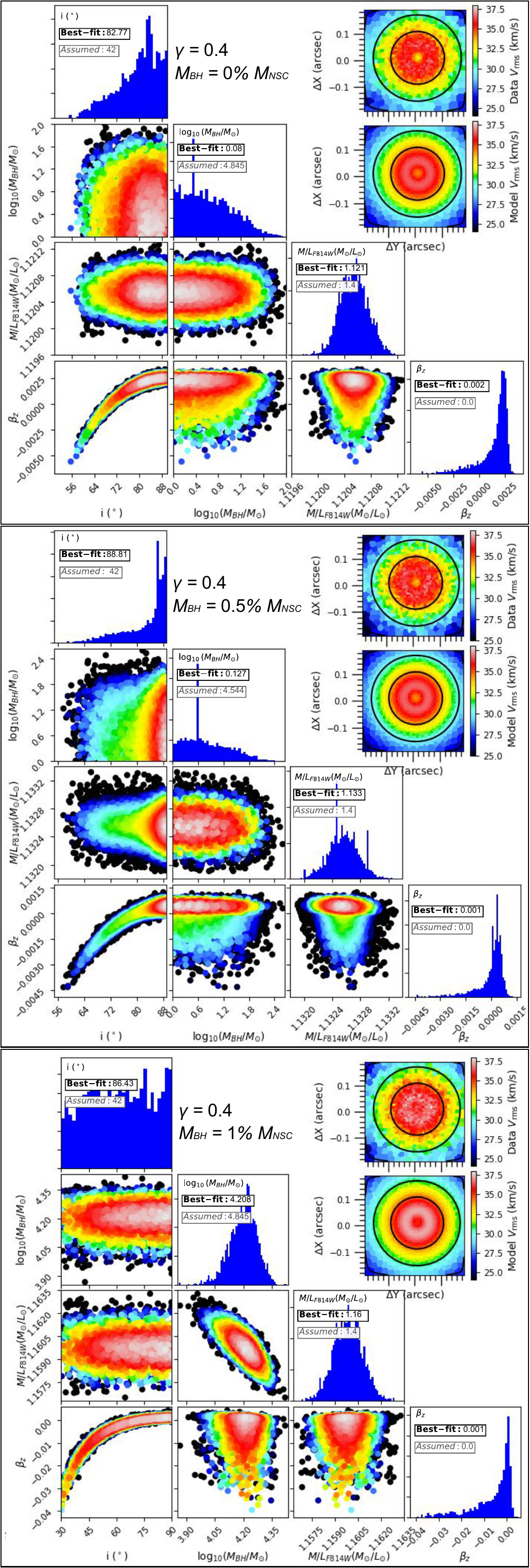}
    \caption{{As} 
 Figure~\ref{new3_ngc3115dw01_gamma_0.0}, but for $\gamma = 0.4$.}
    \label{new3_ngc3115dw01_gamma_0.4}
\end{figure}

\begin{figure}[H]
    
    \includegraphics[width=0.58\textwidth]{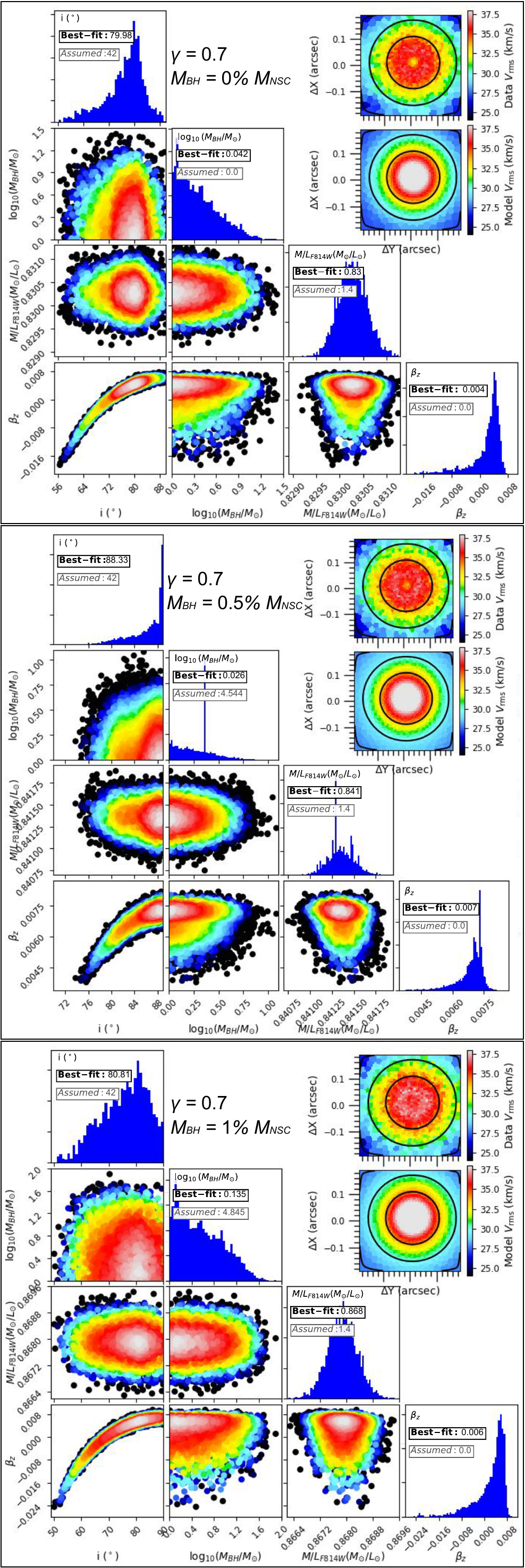}
    \caption{{As} 
 Figure~\ref{new3_ngc3115dw01_gamma_0.0}, but for $\gamma = 0.7$.}
    \label{new3_ngc3115dw01_gamma_0.7}
\end{figure}

\printendnotes[custom] 

\begin{adjustwidth}{-\extralength}{0cm}
\reftitle{References}

\PublishersNote{}
\end{adjustwidth}



\end{document}